\documentclass{article}
\usepackage{graphicx} 
\usepackage{url}
\usepackage{hyperref}
\usepackage{amsmath}
\usepackage{amssymb}
\usepackage{verbatim}
\usepackage{xcolor}
\usepackage{footnote}
\usepackage{physics}
\usepackage{titlesec}
\usepackage{bm}

\titleclass{\subsubsubsection}{straight}[\subsection]

\newcounter{subsubsubsection}[subsubsection]
\renewcommand\thesubsubsubsection{\thesubsubsection.\arabic{subsubsubsection}}

\titleformat{\subsubsubsection}
  {\normalfont\normalsize\bfseries}{\thesubsubsubsection}{1em}{}
\titlespacing*{\subsubsubsection}
{0pt}{3.25ex plus 1ex minus .2ex}{1.5ex plus .2ex}

\makeatletter
\renewcommand\paragraph{\@startsection{paragraph}{5}{\z@}
  {3.25ex \@plus1ex \@minus.2ex}
  {-1em}
  {\normalfont\normalsize\bfseries}}
\renewcommand\subparagraph{\@startsection{subparagraph}{6}{\parindent}
  {3.25ex \@plus1ex \@minus .2ex}
  {-1em}
  {\normalfont\normalsize\bfseries}}
\def\toclevel@subsubsubsection{4}
\def\toclevel@paragraph{5}
\def\toclevel@paragraph{6}
\def\l@subsubsubsection{\@dottedtocline{4}{7em}{4em}}
\def\l@paragraph{\@dottedtocline{5}{10em}{5em}}
\def\l@subparagraph{\@dottedtocline{6}{14em}{6em}}
\makeatother

\title{A Comprehensive Review of TLSNotary Protocol
}
\author{
Maciej Kalka \\
vlayer Labs \\
\texttt{maciej@vlayer.xyz} \\ \\
\and
Marek Kirejczyk \\
vlayer Labs \\
\texttt{marek@vlayer.xyz} \\ \\
}
\date{September 2024}

\begin{document}

\maketitle
\begin{abstract}
    Transport Layer Security (TLS) protocol is a cryptographic protocol designed to secure communication over the internet.
    The TLS protocol has become a fundamental in secure communication, most commonly used for securing web browsing sessions. 
    In this work, we investigate the TLSNotary protocol, which aim to enable the Client to obtain proof of provenance for data from TLS session, while getting as much as possible from the TLS security properties. 
    To achieve such proofs without any Server-side adjustments or permissions, the power of secure multi-party computation (MPC) together with zero knowledge proofs is used to extend the standard TLS Protocol. 
    To make the compliacted landscape of MPC as comprehensible as possible we first introduce the cryptographic primitives required to understand the TLSNotary protocol and go through standard TLS protocol. 
    Finally, we look at the TLSNotary protocol in detail.
\end{abstract}

\newpage
\tableofcontents

\newpage
\section{Introduction}

Transport Layer Security (TLS) protocol has become fundamental in secure communication over the internet. 
It ensures privacy and data integrity between communicating parties, such as web browsers and servers, by encrypting the data transmitted between them. Once data is encrypted and transmitted over a TLS connection, it remains confidential and protected from eavesdropping. 
This means that any sensitive information exchanged between parties, such as usernames, passwords, or financial data, is encrypted during transmission and cannot be intercepted or deciphered by unauthorized entities. 
Additionally, TLS provides authentication, ensuring that the communicating parties are who they claim to be, which prevents man-in-the-middle attacks. 
The protocol also includes mechanisms for data integrity, verifying that the transmitted data has not been altered during transit. 
By combining these features, TLS plays a crucial role in maintaining trust and security in online interactions, making it an indispensable tool for safeguarding sensitive information in various internet applications.

While this level of security and privacy in communication is undoubtedly great, the data that users receive through TLS communication does not have to be signed by the Server.
Consequently, the Client cannot prove the data's origin to any third party, which poses a significant challenge when verifying the authenticity of received information. 

In this work, we explore solutions to the aforementioned challenge of data verifiability in TLS. 
These solutions, known as TLS attestation protocols, are often referred to in the field as Web Proofs or zkTLS. 
The primary goal of these protocols is to produce a verifiably signed attestation of data from a TLS session, either in the form of a signed transcript or a zero-knowledge proof. 
The term "Web Proof" more accurately captures the essence of these protocols, as zero-knowledge proofs are not necessarily a core component of all implementations.
Currently, there are two state-of-the-art approaches to designing such protocols:

\begin{enumerate}
    \item MPC-TLS (Multi-Party Computation TLS)~\cite{Zhang_DECOy2020,TLSNotary_y2021, zkPass_y2023, CDHPR23, XYW23}
    \item Proxy architecture~\cite{GAZBW21, LEFS23, Reclaim_y2023, ELWGS24}
\end{enumerate}

Each of these approaches offers unique advantages and trade-offs in achieving the goal of secure and verifiable data attestation.
This paper focuses on an in-depth examination of one particular MPC-TLS protocol: TLSNotary~\cite{TLSNotary_y2021}. 
We will delve into the technical and cryptographic details of this protocol. 
This approach, leveraging advanced cryptographic tool-set of multi-party computation, allows users to verify the origin and integrity of received data without compromising the confidentiality and security inherent in TLS.

The structure of the whitepaper is following.
It starts with basic cryptographic primitives like Elliptic Curve Diffie-Hellman Key Exchange, Oblivious Transfer, Oblivious Linear Evaluation and associated \texttt{A2M}/\texttt{M2A} protocols and Garbled Circuits, explaining their roles in secure communication.
We then provide an overview of the Transport Layer Security (TLS), focusing on the TLS handshake and AES-GCM mode of encryption.
The core of our paper is dedicated to an in-depth analysis of the TLSNotary protocol. 
We explore its overall concept, potential use cases, and the protocol interface. 
We then dive into the detailed phases of TLSNotary, starting with the MPC-TLS phase, which incorporates a secure three-party TLS handshake and the \texttt{DEAP} (Dual Execution with Asymmetric Privacy) protocol, used for secure two party communication over the TLS connection. 
We also discuss the Notarization phase, which uses a binding commitment scheme, the Selective Disclosure phase for data redaction, and finally, the Data Verification phase. 
Through this comprehensive review, we aim to provide researchers and practitioners with a thorough understanding of TLSNotary's architecture, its security properties, and its potential applications.

\subsection{Notation}
The Web Proofs protocols share a unified conceptual framework. 
The exchange of information occurs among three parties within the TLS protocol, preserving the protocol's original structure. 
To standardize the terminology across various protocols in this work the following terms are used

\begin{itemize}
\item \textit{Client}: The endpoint initiating the TLS connection.
\item \textit{Connection}: A transport-layer connection between two or more endpoints.
\item \textit{Endpoint}: Either the Client or Server or another party participating in the connection.
\item \textit{Handshake}: An initial negotiation between Client and Server that establishes the parameters of their subsequent interactions within TLS.
\item \textit{Notary}: An endpoint who cryptographically signs commitments to the Client's data and the Server's identity
\item \textit{Receiver}: An endpoint that is receiving records.
\item \textit{Sender}: An endpoint that is transmitting records.
\item \textit{Server}: The endpoint that did not initiate the TLS connection.
\item \textit{Verifier}: Application specific party that is not participating directly in the TLS connection
\end{itemize}
Most of these terms are standard designations in the TLS protocol, with the exception of Notary - the third party appearing under different names in the protocols under consideration (TLS Oracle, Verifier, Proxy, Attestor). 
Given the nature of this party's role in the protocols under consideration, we chose the name Notary as best characterizing the function of this endpoint in the procedure.

\newpage
\section{Cryptographic primitives}
In this section we will describe some cryptographic concepts that are fundamental parts of the presented protocols. 
To start with, we recap Elliptic Curve Diffie-Hellman Key Exchange, which is commonly used key exchange technique in TLS protocols.
Then, we include a detailed background on the subparts of multi-party computation schemes: oblivious transfers and garbled circuits.
Description of commitment schemes concludes the section.

\subsection{Elliptic Curve Diffie–Hellman Key Exchange}

The Elliptic Curve Diffie–Hellman (ECDH) Key Exchange operates as an anonymous key agreement scheme, enabling two parties possessing individual elliptic-curve public–private key pairs to establish a shared secret across an unsecured channel. 
It closely resembles the classical Diffie–Hellman Key Exchange (DHKE) algorithm, yet employs elliptic curve scalar multiplication in place of modular exponentiations. 
ECDH relies on the inherent property of EC points

\begin{equation}
\left( a \cdot G \right) \cdot b = \left( b \cdot G \right) \cdot a,
    \label{eq:EC-property}
\end{equation}
where $a,b \in \mathbb{Z}_p$ where $\mathbb{Z}_p$ is the integers mod $p$ field, where $p$ is a prime number, $G$ is an elliptic curve generator point and $a \cdot G$ denotes point multiplication on the elliptic curve.
Suppose the Client and the Server have two secret keys $C_{sk} \in \mathbb{Z}_p$ and $S_{sk} \in \mathbb{Z}_p$ respectively. 
The 2-party ECDH key exchange works as follows:
\begin{enumerate}
    \item Client generates a public key $C_{pk} = C_{sk} \cdot G$
    \item Server generates a public key $S_{pk} = S_{sk} \cdot G$
    \item Client and Server exchange their public keys
    \item Client calculates shared secret $k = S_{pk} \cdot C_{sk}$
    \item Server calculates shared secret $k = C_{pk} \cdot S_{sk}$
\end{enumerate}
As the condition (\ref{eq:EC-property}) holds the shared keys are the same.
The security of the protocol is relied on the elliptic curve discrete logarithm problem.

\subsection{Oblivious Transfer}

\begin{figure}[ht]
    \centering
    \includegraphics[width = 0.75\linewidth]{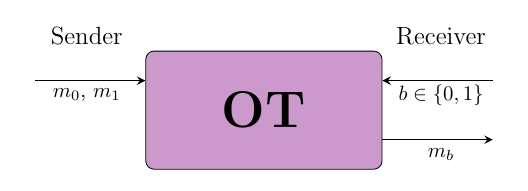}
    \caption{Schematic representation of OT}
    \label{fig:OT}
\end{figure}

One of the fundamental problems in multi party computation is sending one of many pieces of information (e.g. data or message) in such a way that a sender does not know which piece of information was read by receiver.
As the sender remains oblivious to which part of the information was transmitted, such transfer is called an oblivious transfer (OT) and was introduced in the seminal work of Rabin~\cite{Rabin1981}.
Simple form of the OT is a two party protocol that gets two input bits $m_0$, $m_1$ (called later \textit{messages}) from a sender and one input bit $b \in \{0,1\}$ from a receiver, and then returns $m_b$ to the receiver. 
Upon completion of the protocol, the receiver is expected to acquire solely the message $m_b$ and nothing more, while the sender is meant to gain no knowledge about receiver's choice.
Such setup, known as 1 out of 2 OT has been shown in the Fig.~\ref{fig:OT} and can be easily extended to 1 out of N variant.

\noindent
As an illustrative example we will present a simple 1 out of 2 OT based on the ECDH key exchange.
Suppose that the Client has two plain text messages $m_0^C$ and $m_1^C$ and the Notary has a bit $b \in \{0,1\}$ indicating he wants to read message $m_b^C$ without Client knowing which message he has read.
The OT protocol based on ECDH then reads as follows

\begin{enumerate}
    \item Client and Notary generate secret keys $C_{sk}, N_{sk} \in \mathbb{Z}_p$ respectively
    \item Client generates a public key $C_{pk} = C_{sk} \cdot G$ and shares it with the Notary
    \item Notary generates a public key depending on which message he wants to read 
    \begin{equation}
    \left\{
        \begin{array}{llll}
           N_{pk}  &=& N_{sk} \cdot G & \text{ if } b = 0  \\
           N_{pk}  &=& C_{pk} + N_{sk} \cdot G & \text{ if } b = 1 
        \end{array}
    \right.
    \end{equation}
    \item Notary generates a secret $k_b = C_{pk} \cdot N_{sk} = (C_{sk} \cdot G) \cdot N_{sk}$
    \item Client generates two secrets
        \begin{equation}
    \left\{
        \begin{array}{lll}
           k_0  &=& C_{sk} \cdot N_{pk}  \\
           k_1  &=& C_{sk} \cdot N_{pk} - C_{sk} \cdot C_{pk} 
        \end{array}
    \right.
    \end{equation}
    \item Client shares encrypted messages $\texttt{Enc}(k_0, m_0)$ and $\texttt{Enc}(k_1, m_1)$ with the Notary
    \item Notary is only able to decrypt one of the ciphers since $k_b$ is identical either with $k_0$ or $k_1$
\end{enumerate}
Following this procedure the secret key $k_b$ forged by the Notary enables him to decrypt only one message depending on his choice of $b$ without revealing his choice to Client.
The protocol has been introduced and tested for security by~\cite{Chou_2015}.

\subsubsection{Reduced Oblivious Transfer techniques}
\begin{figure}[ht]
    \centering
    \includegraphics[width = 0.48\linewidth]{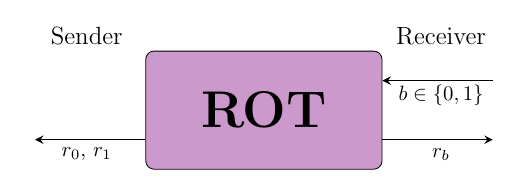}
    \includegraphics[width = 0.48\linewidth]{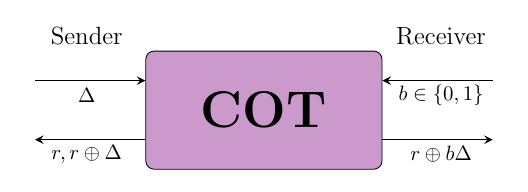}
    \caption{Schematic representations of ROT and COT}
    \label{fig:OT-red}
\end{figure}

So far we have discussed the most general case of the OT where Sender had a set of input strings and receiver chose one of them using his input selection index.
Now we will look at two simplified protocols - Random and Correlated OT, where the Receiver still inputs his selection index, but the inputs of the Sender are significantly reduced.
Schematic representation of these protocols are shown in the Fig.~\ref{fig:OT-red}

In the 1 out of 2 Random OT (ROT) setting the receiver inputs a selection index, and the protocol outputs set of 2 pseudo-random $l$-bit strings to the Sender and one string to the Receiver selected according to his index.
This idea simply extends into 1 out of $N$ ROT.
On the other hand let us consider an OT where the Sender's inputs are correlated with each other, but at the same time they are completely random.
Such functionality is achieved with Correlated OT (COT).
In the 1 out of 2 setting the Sender inputs a fixed constant $\Delta$ and the Receiver inputs his selection index.
As an output Receiver gets two random messages $r_0, r_1$, which are correlated by $\Delta$.
This correlation can be expressed by two messages xoring up to a $\Delta$ namely $r_0 \oplus r_1 = \Delta$.
Consequently, we can rephrase the messages and denote them as $r, r\oplus\Delta$ as showed in the Fig.~\ref{fig:OT-red}.
In such setup, Receiver outputs an $r_b$, which in terms of $\Delta$ is given by $r\oplus\Delta b$.

{
\subsection{Oblivious Linear Evaluation}

Following the discussion on Oblivious Transfer (OT), where one party securely transfers one of many pieces of information without knowing which was chosen, we delve into Oblivious Linear Evaluation (OLE). 

\begin{figure}[ht]
    \centering
    \includegraphics[width = 0.75\linewidth]{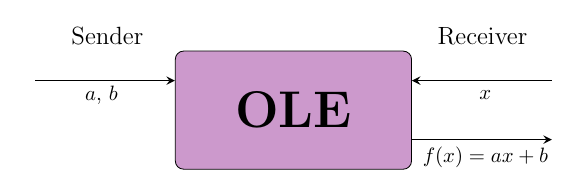}
    \caption{Schematic representation of OLE}
    \label{fig:OLE}
\end{figure}
\noindent
OLE builds on OT by allowing one party to compute a linear function using inputs from another party, without revealing anything beyond the result. This makes OLE an essential protocol in secure multi-party computation.
A schematic representation of OLE functionality is presented in Fig~\ref{fig:OLE}.
As in OT there are two parties -- Sender and Receiver.
Sender provides two private finite field elements $a, b \in \mathbb{F}_p$ while Receiver provides his private input $x \in \mathbb{F}_p$.
As a result of a OLE functionality Receiver gets a value of linear polynomial $f(x) = ax + b$.
One should note that that way OLE can be seen as a generalization of OT.
With $a = m_0$ and $b = m_1 - m_0$ OLE implements OT.
The linear function $f(x)$ reads as follows $f(x) = (m_1 - m_0)x + m_0$, where Receiver input $x = \{0, 1\}$ represents a bit indicating message $m_x$ that Receiver wants to read.

A potential security weakness in OLE is the fact that Receiver can exploit the process by setting the his input $x = 0$.
When this happens, the computed output simplifies to the value $b$, effectively revealing the constant term selected by the Sender.
In order to use OLE as a building block of a secure multi-party computation one has to address that functionality issue.
If we wanted to use OLE as secret-shared multiplication a simple way to solve this security problem would be selecting $b$ at random.
That way even if malicious Receiver wants to trick the protocol by setting $x=0$ he doesn't learn Sender's secret input, but just a random number.
Sender inputs his secret $a$ and random $b$ and Receiver inputs his secret $x$.
Simple algebraic manipulation on the first degree polynomial leads to expression $y - b = ax$.
Such relaxation enables us to view OLE as a secret-shared multiplication.

\subsubsection{Variants of OLE}

Two particularly significant variants of Oblivious Linear Evaluation are Random OLE (ROLE) and Vector OLE (VOLE), which extend the capabilities of basic OLE in distinct ways. Figure~\ref{fig:OLE-var} provides a schematic representation of these two protocols, illustrating their key components and operations. 

\begin{figure}[ht]
    \centering
    \includegraphics[width = 0.48\linewidth]{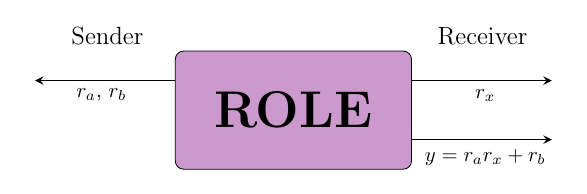}
    \includegraphics[width = 0.48\linewidth]{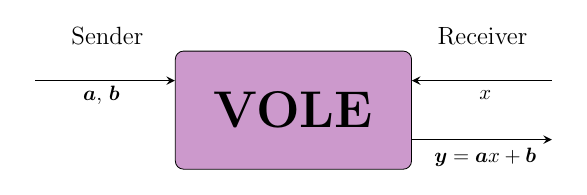}
    \caption{Schematic representations of ROLE and VOLE}
    \label{fig:OLE-var}
\end{figure}

In ROLE the sender inputs random values $r_a$ and $r_b$, while the receiver inputs a random value $r_x$. 
The protocol then computes $y = r_ar_x + r_b$, with the Receiver obtaining this result. 
This randomized approach allows for the generation of correlated randomness between parties without revealing individual inputs, making ROLE a valuable building block for more complex cryptographic MPC schemes.
VOLE, depicted on the right side of Figure 4, extends the OLE concept to vectors. 
In this case, the sender inputs private vectors $\bm{a}$ and $\bm{b}$, while the receiver inputs a private scalar value $x$. 
The protocol computes $\bm{y} = \bm{a}x + \bm{b}$ elementwise, resulting in a vector output. 
This vectorized operation is particularly useful for scenarios involving large-scale data and repeated operations as one instance of VOLE performed with $n$-dimensional vectors is equivalent to performing OLE $n$ times.
}
\subsection{\texttt{M2A} and \texttt{A2M} protocols}

Fundamentally the concept of the secure two-party computation boils down to the problem of computing the pre-agreed function $f(x_C, x_N)$ where $x_C$ and $x_N$ are secret inputs belonging to the Client and the Notary respectively without revealing any information about the inputs.
In general that problem can be securely solved within garbled circuits framework, although as the implementation complexity of the garbled circuit strongly relies on boolean circuit complexity of the function $f$, more efficient protocols for the basic arithmetic operations has been proposed.
In the paper on such protocol for efficient two-party exponentiation Yu et al. introduce two sub-protocolos: \texttt{A2M} and \texttt{M2A} which are particularly relevant in other applications which involve converting two-party sharings of sums into products and vice versa~\cite{Yu_2011}.
The interface for the protocol \texttt{A2M} (additive sharing to multiplicative sharing) reads as follows

\vspace{10pt}
\noindent\fbox{%
    \parbox{\textwidth}{%
\noindent
\texttt{A2M} protocol interface
    
\noindent
\textbf{Inputs:} Client holds a secret $x_C \in \mathbb{Z}_p$, Notary holds a secret $x_N \in \mathbb{Z}_p$ and $x = x_C + x_N \in \mathbb{Z}^{\ast}_p$

\noindent
\textbf{Outputs:} Client obtains a secret $z_C$ and Notary obtains a secret $z_N$ such that $z_C \cdot z_N = x$
}
}
\vspace{10pt}

\noindent
Consequently the interface for the protocol \texttt{M2A} (multiplicative sharing to additive sharing) is given by the following scheme

\vspace{10pt}
\noindent\fbox{%
    \parbox{\textwidth}{%
\noindent
\texttt{M2A} protocol interface

\noindent
\textbf{Inputs:} Client holds a secret $x_C \in \mathbb{Z}_p^{\ast}$, Notary holds a secret $x_N \in \mathbb{Z}_p^{\ast}$ and $x = x_C \cdot x_N \in \mathbb{Z}^{\ast}_p$

\noindent
\textbf{Outputs:} Client obtains a secret $z_C$ and Notary obtains a secret $z_N$ such that $z_C + z_N = x$
}
}

\noindent
\vspace{10pt}

\noindent
{
Above mentioned protocols can be implemented in maliciously secure way using oblivious transfer~\cite{Yu_2011}.

\subsubsection{\texttt{M2A} using OLE}

Knowing the definition of Oblivious Linear Evaluation it seems appealing to use it to convert shares of a product into shares of a sum.
To do so, let us define OLE-M2A protocol in the following way:

\vspace{10pt}
\noindent\fbox{%
    \parbox{\textwidth}{%
\noindent
OLE based \texttt{M2A} protocol interface

\noindent
\textbf{Inputs:} Client holds a secret $x_C$, Notary holds a secret $x_N$

\noindent
\textbf{Outputs:} Notary obtains a secret $z_N$ and Client obtains a secret $z_C$ such that $z_C = x_Cx_N + z_N$
}
}

\noindent
\vspace{10pt}

\noindent
After getting input $x_C$ from Client and $x_N$ from Notary ordinary OLE functionality returns $z_N$ to Notary and $z_C$ to Client such that $z_C = x_Cx_N + z_N$.
Problems arise in the presence of a malicious adversary, who inputs $0$ into the protocol, because now $z_C = z_N$, which means that privacy for the honest party's output is no longer guaranteed.

To address this shortcoming both parties can run the following protocol, which allows an honest party to detect an input of $0$. 
Let's assume Notary is honest and wants to check if Client used a 0 input. 

\begin{enumerate}
    \item Client chooses OLE input $a_k$ and Notary chooses $x_k$ for $k = \{1, \dots l\}$. These are the base OLEs we want to check for 0 input of Client.
    \item Both parties run Random OLE functionality to generate $(a_0, x_0)$ for Client and $(b_0, y_0)$ for Notary such that $y_0 = a_0x_0 + b_0$. This is needed as a mask later.
    \item Client sets $a_{k + 1} = a_{k - l}^{-1}$ , and Notary sets $x_{k + 1} = x_{k - l}$ for $k = \{l,\dots,2l\}$.
    \item Both parties call OLE with their inputs from (1) $a_k$ and $x_k$. OLE returns secret $b_k$ and $y_k$ such that $y_k = a_kx_k + b_k$ for $k = \{1,\dots,2l + 1\}$. 
    \item Client computes $ \displaystyle s = \sum_{k = 0}^l - x_k \cdot a_k^{-1} + \sum_{k = l + 1}^{2l +1} - x_k$
    \item Client sends $s$ to Notary.
    \item Notary checks that $\displaystyle \sum_{k = 0}^l b_k  = s + \sum_{k = l + 1}^{2l + 1} y_k $. If this does not hold Notary aborts.
\end{enumerate}
The protocol can be repeated with roles swapped, to also check that Notary was honest.

The description provided above only intuitively sketches the solution to zero-check problem in OLE based \texttt{M2A} protocol.
However, both \texttt{M2A} and \texttt{A2M} protocols can be implemented formally in secure way using oblivious linear evaluation~\cite{KOS16, XYW23}.
}

\subsection{Garbled Circuits}
\begin{figure}[ht]
    \centering
    \includegraphics[width=\linewidth]{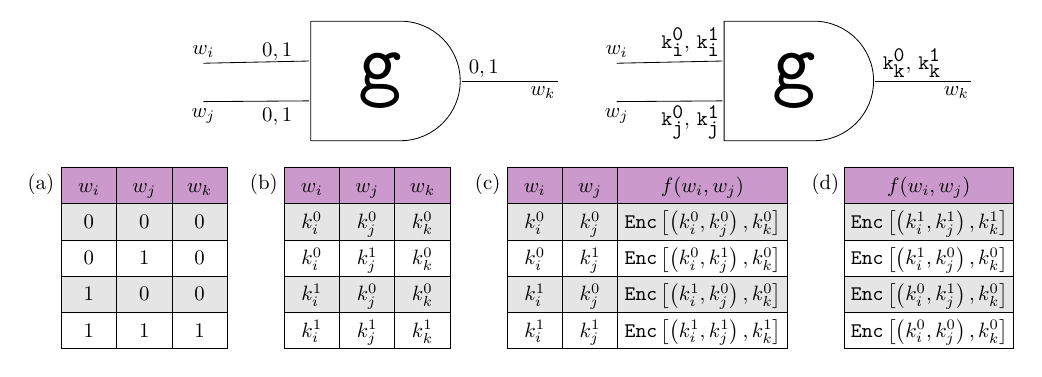}
    \caption{Garbling scheme for AND gate: (a) AND gate before encryption, (b) gate encoded with random input and output labels, (c) output label encrypted with symmetric keys derived from input labels, (d) encrypted gate: a randomly permuted (garbled) output label table}
    \label{fig:GC}
\end{figure}

The history of multi-party computation starts with the garbling technique shown by Yao~\cite{Yao_1986}, which has been later shown computationally secure.
Let us consider two parties, Client and the Notary, who want to compute a function $y = f(x_C, x_N)$ in such a way that Client wants to learn $y$ and keep secret his input $x_C$ and the Notary wants to learn $y$ without revealing his secret input $x_N$.
We start with writing the function $f$ as a boolean circuit, using logic gates.
Each logic gate in the system has two inputs and one output, so called wires which can have two values on it i.e. $w_i \in \{0, 1\}$.
Then each gate \texttt{g} is described by two input wires and one output wire namely $\texttt{g}(w_i, w_j) = w_k$.
Once the function is converted to boolean circuit and both parties know it, we need to assign the roles of Garbler and Evaluator among Client and Notary. 
As an illustrative example let us go through the garbling scheme for AND gate which is presented in Figure~\ref{fig:GC}.
The truth table for AND gate is shown in the Fig.~\ref{fig:GC}(a).
The cryptographic "encryption" of the logic gate starts by encoding the wires.
Intuitively speaking, the encoding of a wire is basically assigning random value (called wire label) to each wire value.
Namely for the wire $w_i$ the labels are $k_i^0$ and $k_i^1$.
The correspond to $w_i = 0$ and to $w_i = 1$ respectively.
Then the random labels can be used for output encryption.
The encrypted output which is the value of the function $f$ can be written using input labels as two symmetric keys encrypting the output label, namely
\begin{equation}
    f(k_i^a, k_j^b) = \texttt{Enc}\left[(k_i^a, k_j^b), k_k^c \right]
\end{equation}
where $a, b, c \in \{0,1\}$ and the two keys are the input labels.
That procedure repeated for every combination of input labels in the truth table of the gate results in encrypted truth table for the given logic gate.
Finally the order of the encrypted values of $f$ in the encrypted truth table is garbled, which means randomly permuted.
That way that given a garbled truth table of the function $f$ the output values cannot be deduced.

\noindent
Now we will present the general interface for garbling protocol for the secure function evaluation.
For ease of understanding in the description below we will refer to the presented garbling scheme for single AND gate.
Note that in order to apply this protocol for general case of any circuit you need to repeat the garbling and evaluation procedure for each gate in the circuit.
Garbling protocol steps for secure function evaluation between Client and the Notary read as follows
\begin{enumerate}
    \item We assume that Client is the Garbler and Notary is the Evaluator.
    Client creates a garbled circuit for the function $f(x_C, x_N) = v$. 
    Garbled circuit for the function $f$ is given by the tuple $(G, (I_C, I_N), O)$:
        \begin{itemize}
            \item $G$ is a set of gates representing a function $f$
            \item $I_C$ is the input wire label table for Client's private variable (Fig 3b, column 1)
            \item $I_N$ is the input wire label table for Notarys's private variable (Fig 3b, column 2)
            \item $O$ is the output wire label translation table (Fig 3b, column 3 with corresponding outputs)
            \begin{table}[ht!]
                \centering
                \begin{tabular}{|c|c|}
                    \hline
                    label & output \\ \hline
                    $k_0^k$ & 0 \\ \hline
                    $k_1^k$ & 1 \\ \hline
                \end{tabular}
            \end{table}
        \end{itemize}
    \item Client sends Notary: 
        \begin{itemize}
            \item The garbled gates with output labels encrypted with symmetric key encryption scheme (Fig 3d).
            The input wire labels are used as symmetric keys
            \item His encoded active input $x_C$. 
            It is the one, selected element of his input wire label table $I_C$ e.g. $k_i^0$
            \item Output wire label translation table $O$ which is two random strings with corresponding output values
        \end{itemize}
    \item For the evaluation, Notary has to use his encoded input $x_N$ as well. 
    As only the Client knows how to garble the circuit, parties execute oblivious transfer protocol.
    Notary is the receiver in this OT -- he inputs a bit $b$ corresponding to the output he want to receive from the Client.
    Using OT Client sends Notary only an element of $I_N$ corresponding to Notary's bit $b$ without knowing which bit Notary have chosen.
    \item Notary can now evaluate garbled circuit and get the encrypted output wire labels.
    \item Importantly, as the outputs are encrypted using symmetric keys Notary can decrypt only one output -- the one corresponding to the bit he choose during OT in step 3.
    \item Using output wire label translation table $O$ the output is decoded by Notary.
    \item The Notary (Evaluator) sends the decrypted and decoded output to Client (Garbler)
\end{enumerate}
An important security property of garbled circuits is that given a garbled circuit, garbled input and decoding information reveals nothing beyond function $f(x_C, x_N)$. 

\subsubsection{Garbling scheme formalization}
The garbling procedure described above in an intuitive way has been formalized in the work of Bellare et al.~\cite{Bellare_2012}.
In this section we would like to briefly summarize this formalization, as the notions used by Bellare et al. frequently appear in abstract descriptions of the protocols which use garbling schemes.
A garbling scheme $\mathcal{G}$ consists of five algorithms $(\texttt{Gb}, \texttt{En}, \texttt{Ev}, \texttt{De}, \texttt{ev})$.
At first, the function $f$ is transformed to a circuit.
The algorithm $\texttt{ev}$ allows to evaluate the circuit $\texttt{ev}(f, x) = f(x)$.
The circuit $f$ becomes an input to the randomized garbling algorithm $\texttt{Gb}$.
The algorithm works the following way
\begin{equation}
    \texttt{Gb}(1^k, f) = (F, e, d)
\end{equation}
returning a three-tuple of strings which are interpreted as functions: $F$ -- the garbled circuit, $e$ -- the encoding information and $d$ the decoding information.
Then the the encoding information $e$ is passed to the $\texttt{En}$ algorithm (\texttt{Encode}) which produces a garbled input $X$ from the initial input $x$
\begin{equation}
    \texttt{En}(e, x) = X.
\end{equation}
This step correspond to the input wires encoding described in previous section.
In the abstract fashion the encryption information $e$ can be an arbitrary string, although is we require the garbling scheme to be projective then $e$ consists of $2n$ wire labels, where $n$ is the number of input bits.\footnote{As discussed by Bellare et al the projectivity of garbling scheme is important in several applications (e.g. Secure Function Evaluation). For more details see~\cite{Bellare_2012}}
Then encoding in the projective garbling scheme directly relates to the example of encoding AND gate presented earlier.
Having one input bit on each of two the wires $w_1 = \{0, 1\}$ and $w_2= \{0, 1\}$ the encoding information consists of wire labels which are assigned to the wires within encoding algorithm 
\begin{equation}
 \texttt{En}\left(e = \{k^0_i, k^1_i\}_{i=1,2}, x = (w_1, w_2)\right)  = \{k^{0}_{w_1}, k^{1}_{w_1}, k^{0}_{w_2}, k^{1}_{w_2}\}
\end{equation}
Third step of garbling scheme is the $\texttt{Ev}$ algorithim (\texttt{Evaluate}) which takes the garbled circuit $F$ together with garbled input $X$ and returns a garbled output $Y$
\begin{equation}
    \texttt{Ev}(F, X) = Y
\end{equation}
Finally the decoding information $d$ is used in $\texttt{De}$ algorithm (\texttt{Decode}) to  decode the garbled output and acquire the final output
\begin{equation}
    \texttt{De}(d, Y) = y
\end{equation}
The corectness condition for the garbling scheme is that final output $y$ is equal to the $f(x)$ and formalizes as follows
\begin{equation}
    \texttt{De}(d, \texttt{Ev}(F, \texttt{En}(e, x))) = \texttt{ev}(f,x)
\end{equation}

\subsubsection{Nontrivial example}

\begin{figure}[ht!]
    \centering
    \includegraphics[width=0.75\textwidth]{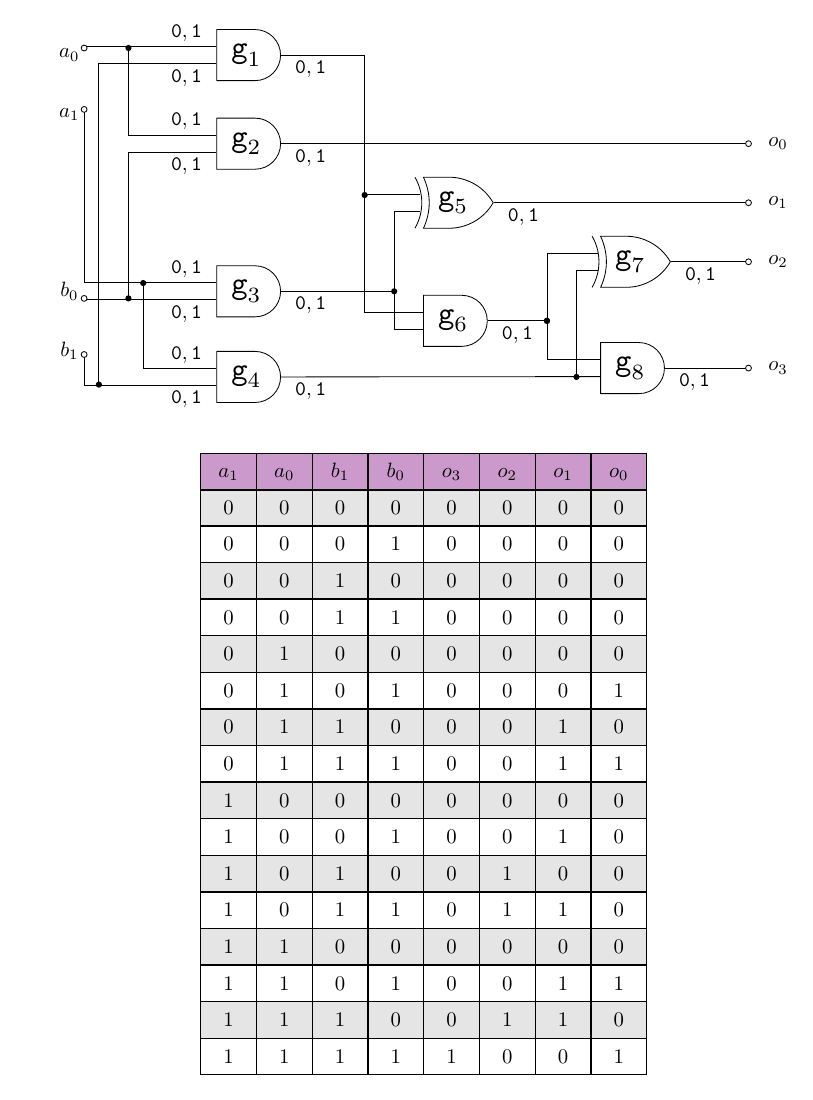}
    \caption{Logic gate circuit for 2 bit by 2 bit multiplier function with truth table.}
    \label{fig:GC-example-Decrypted}
\end{figure}

Let us follow the garbling scheme for the 2 bit by 2 bit binary multiplication.
The function which is nontrivial yet simple enough to comprehend.
We assume that Client owns 2 bit binary number $A = a_1a_0$ and Notary owns the 2 bit binary number $B=b_1b_0$ and they want to evaluate the function $f(A, B) = AB$.
For the considered function the circuit consists of 8 gates -- 6 AND gates and 2 XOR gates.
It is shown in Fig.~\ref{fig:GC-example-Decrypted}.

\begin{figure}[ht!]
    \centering
    \includegraphics[width=0.85\textwidth]{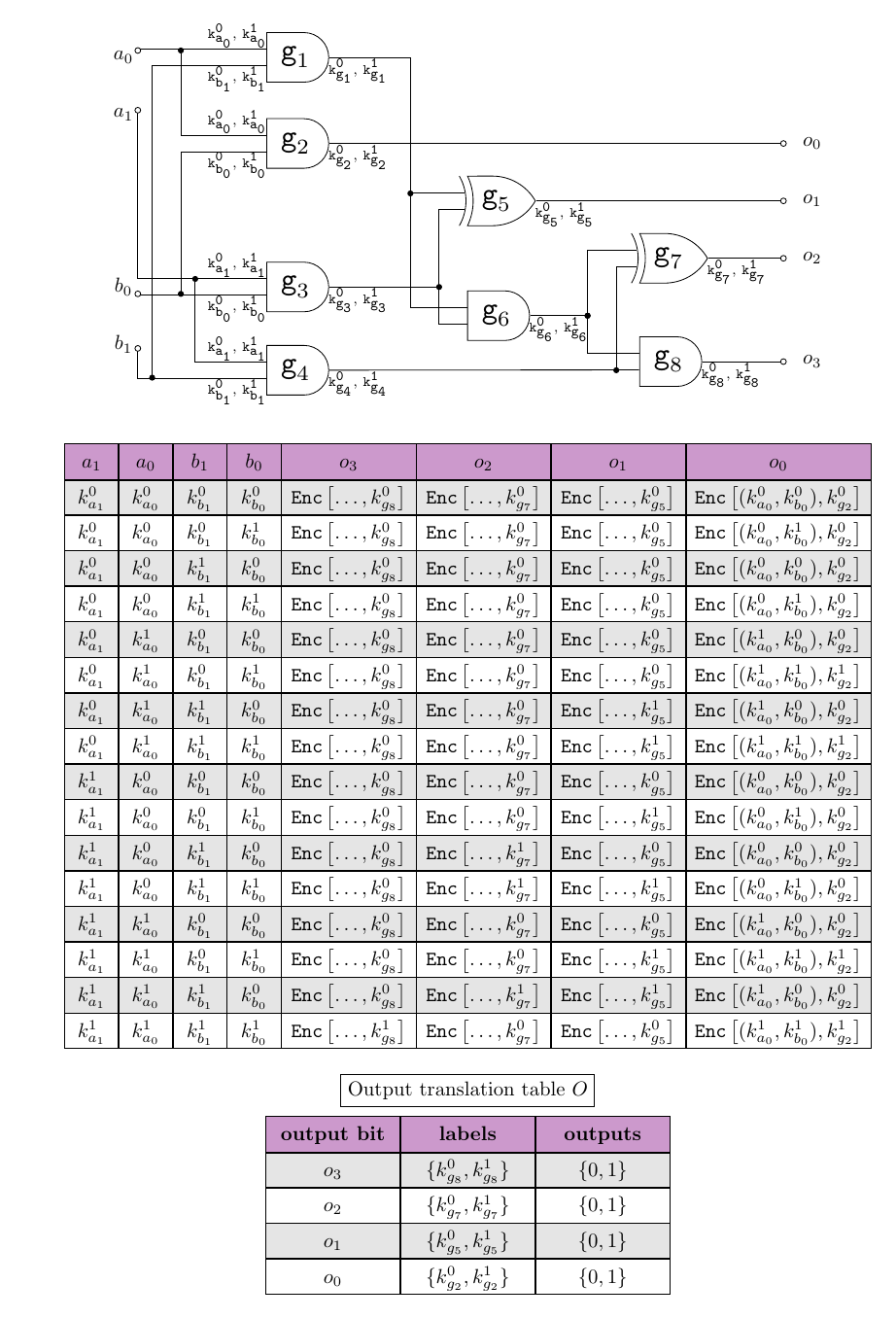}
    \caption{Logic gate circuit for 2 bit by 2 bit multiplier function with encoded input and output wire labels. Below the encrypted truth table for the circuit is presented. For the sake of readability the full encryption scheme with both keys is presented only for output bit $o_0$. The encryption keys for remaining output bits can be found by following the circuit. The output translation table $O$ is presented below the truth table. It is sent from Garbler to Evaluator in the second step of the secure function evaluation with garbling scheme protocol.}
    \label{fig:GC-example}
\end{figure}

The garbling scheme starts with the Garbler (Client) rewriting the function $f$ as a logical gate circuit.
Then, Garbler encrypts each gate in the circuit.
He picks random wire labels for all of the input and output wires in the circuit.
For the simplicity of notation in Fig.~\ref{fig:GC-example} we assign the symbols $g_i$ to the consecutive gates. 
Then we describe the wire labels as $k^i_j$, where $i$ is either 0 or 1 and $j$ refers to the symbol assigned to the input bit $(a_0, a_1, b_0, b_1)$ in case of input wires and in the case of output wires $j$ refers to the symbol assigned to the gate ($g_1, g_2, \dots, g_8$).
Then the Garbler encrypts a truth table for each gate in the circuit, which results in encryption of every output wire label (including final output bits $o_0, \dots, o_3$).
Garbler prepares output translation table $O$ (see Fig.~\ref{fig:GC-example}) and the setup phase is finished.
The Garbler sends a garbled circuit (set of encrypted randomly shuffled gates) to the Evaluator together with input labels corresponding to his selected input to the Evaluator.
This concludes the second step of the protocol.
For the the step three of the protocol, the Notary (the Evaluator) requests input labels corresponding to his desired input using Oblivious Transfer.
That enables Evaluator to calculate only one output per gate, which corresponds only to his labels of choice.
Evaluator gets the encrypted wire labels
If you know the labels of both inputs, and you have the garbling, then you can learn the label of the corresponding output and after decryption divide it out to learn other party input.
This makes multiplication fairly pointless use of garbled circuit as the minimal functionality (reversibility) of multiplication just breaks the privacy.
In general garbled circuits can be used for computations that are not reversible, and so don't break privacy in this way

\subsubsection{Garbling circuit optimizations}
As demonstrated in the example of multiplication, garbled circuits are characterized by high complexity, both in terms of the number of logic gates that constitute the function and the number of labels that must be assigned to the gate wires. 
This complexity arises from the need to encrypt and manage a vast amount of data securely, which can be computationally intensive and resource-demanding. 
Consequently, numerous optimizations of garbled circuits have been developed over the years to enhance the efficiency and practicality of secure multi-party computation (MPC) using this primitive. 
These optimizations aim to reduce the computational overhead, minimize the communication costs, and improve the overall performance of garbled circuits. Techniques such as free XOR gates, row reduction, and half-gates have been introduced to streamline the process and make secure MPC more feasible for real-world applications.

Readers interested in detailed descriptions of these optimization techniques and comparisons of their efficiency can refer to seminal works in the field. For instance, the free XOR technique significantly reduces the number of encrypted gates required for certain operations~\cite{Kolesnikov2008}. 
The row reduction method optimizes the storage and transmission requirements of garbled tables~\cite{Pinkas2009}.
Additionally, the half-gates approach provides an efficient alternative to traditional garbled circuit constructions, reducing both the computation and communication complexities~\cite{Zahur2014}. 
These studies offer comprehensive insights into the advancements and practical implementations of garbled circuit optimizations.

\subsection{Commitment Schemes}
Commitment scheme is a protocol used whenever one party want to commit to a certain message or value without revealing it immediately, while ensuring they cannot change the committed value later.
The protocol is particularly useful in secure communication or scenarios like voting or auction systems.
The commitment scheme interface between Client and Notary, where Client is the committer and Notary is the receiver is the following

\vspace{10pt}
\noindent\fbox{
    \parbox{\textwidth}{
\noindent
Commitment scheme interface
    
\noindent
\textbf{Inputs:} Client's original value $x_C$, ${C()}$ is the commitment function and ${V()}$ is the verification function

\noindent
\textbf{Outputs:} Commitment to value $x_C$ $\texttt{com}_{x_C} = C(x_C)$, validation result of function $V$ (boolean)
}
}
\vspace{10pt}

\noindent
The protocol consists of three steps
\begin{enumerate}
    \item Client selects the value to be committed $x_C$ and generates $\texttt{com}_{x_C}$ using function $C()$
    \item When Client is ready he reveals his value $x_C$ and the commitment scheme $\texttt{com}_{x_C}$
    \item Notary verifies if revealed value matches the revealed commitment.
    If so -- the Client did not change the value he committed to.
\end{enumerate}
The commitment function $C(x_C)$ should be at the same time be computationally binding and hiding.
Meaning it should be effectively impossible to change the committed value without detection and given only $C(x_C)$ it should be computationally infeasible for anyone to determine what the original value $x_C$ is.
A simple example of commitment scheme can be presented using SHA-256 hash function to create a commitment to a given value. 
Verification function validates is SHA-256 hash of a given value equals to the presented commitment.

\newpage
\section{Introduction to Transport Layer Security}

Transport Layer Security (TLS) is a cryptographic protocol designed to secure communication over networks, most commonly used for securing web browsing sessions. 
It has evolved from its predecessor, the Secure Sockets Layer (SSL).
While SSL 3.0 was the first widely deployed version but was later found to have significant vulnerabilities. 
TLS 1.0 was introduced in 1999 as an upgrade to SSL 3.0, addressing some of its security flaws. 
It was followed by TLS 1.1 in 2006, offering further improvements. 
However, it wasn't until TLS 1.2, released in 2008, that significant strides were made in security. 
TLS 1.2 introduced stronger cryptographic algorithms, enhanced the handshake process, and mitigated known attacks. 
It is widely regarded as secure and is the recommended standard for secure communication on the internet.
The latest version, TLS 1.3, released in 2018, builds upon TLS 1.2 by further improving security and performance, reducing handshake latency, and removing outdated cryptographic algorithms. 
TLS provides privacy and data integrity between communicating parties by encrypting the data transmitted between them.
A secure connection between a Client and a Server is established, ensuring that data exchanged between them cannot be intercepted or tampered with by malicious third parties.
As stated in the original documentation of TLS 1.3~\cite{Rescrola_TLS13y2018}

\noindent
"The primary goal of TLS is to provide a secure channel between two communicating peers; the only requirement from the underlying transport is a reliable, in-order data stream. Specifically, the secure channel should provide the following properties:
\begin{itemize}
    \item \textbf{Authentication}: The Server side of the channel is always authenticated; the Client side is optionally authenticated. Authentication can happen via asymmetric cryptography (e.g., RSA, the Elliptic Curve Digital Signature Algorithm (ECDSA)), or the Edwards-Curve Digital Signature Algorithm (EdDSA) or a symmetric pre-shared key (PSK).
    \item \textbf{Confidentiality}: Data sent over the channel after establishment is only visible to the endpoints. TLS does not hide the length of the data it transmits, though endpoints are able to pad TLS records in order to obscure lengths and improve protection against traffic analysis techniques.
    \item \textbf{Integrity}: Data sent over the channel after establishment cannot be modified by attackers without detection."
\end{itemize}

\subsection{TLS Handshake}
\begin{figure}[ht!]
    \centering  \includegraphics[width=0.7\linewidth]{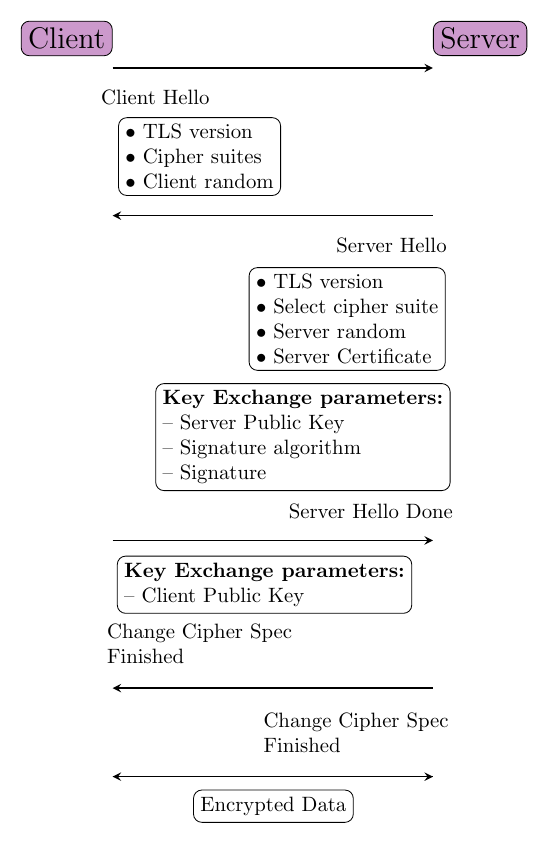}
    \caption{Message flow in the TLS 1.2 handshake}
    \label{fig:TLS-12}
\end{figure}

The TLS protocol has two main components: handshake and record protocols.
To understand the differences between TLS 1.2 and 1.3 let us take a look at the handshake protocol in TLS 1.2 first.
In the Fig~\ref{fig:TLS-12}
the message flow of TLS 1.2 handshake is presented

\noindent
In case of TLS 1.2 handshake communication consists of two round trips (RT) between Client and Server.
The protocol starts with Client Hello message.
Client tells the Server what TLS version he want to use, list of cipher suites available for him and a random 28-byte number (Client Random).
That is followed by Server Hello message, where Server checks if TLS version is valid, picks preferred cipher suite from Client's list and adds Server certificate to prove identity.
Once key exchange algorithm is established (after selecting cipher suite) key exchange parameters are generated and put to the message.
Last but not least, before sending Server Hello message, Server saves Client Random and generates its own Server Random 28-byte number.
In the next step Client verifies Server certificate and signature in key exchange algorithm parameters.
When checks are complete Client adds his public key to the key exchange algorithm parameters.
Pre-Master Secret (PMS) is generated using Server's and Client's key exchange parameters.
Later a Master Secret (MS) is created independently by Server and Client out of their Randoms and PMS.
The session keys for communication are generated based on the Master Secret.
Once MS is ready Client sends Change Cipher Spec message letting Server know that all future messages are encrypted.
Client's Finished message includes the whole message flow but in cipher text.
This way Server can verify if they agreed on the same encryption, preventing man in the middle attack.
To finish second RT Server sends similar messages and finishes with cipher text version of previous messages for encryption verification.
Once the handshake is done Server and Client are communicating in HTTPS.

The key changes introduced the TLS 1.3 relate to improvements in speed and security of the protocol.
Security-wise, the support for legacy non-AEAD ciphers has been removed. 
TLS 1.3 simplified cipher suites by removing separate key exchange algorithms, as it always uses Diffie-Hellman key exchange.
Thus, the list of supported cipher suites has been shorten to 5 (compared to 37 in TLS 1.2) namely
\begin{enumerate}
    \item \texttt{TLS\_AES\_128\_GCM\_SHA256}
    \item \texttt{TLS\_AES\_256\_GCM\_SHA384}
    \item \texttt{TLS\_CHACHA20\_POLY1305\_SHA256}
    \item \texttt{TLS\_AES\_128\_CCM\_SHA256}
    \item \texttt{TLS\_AES\_128\_CCM\_8\_SHA256}
\end{enumerate}
The first three cipher suites are mandatory for all TLS 1.3 implementations, while the latter two (CCM-based) are optional.
Each cipher suite designation encapsulates multiple cryptographic elements:

\begin{itemize}
    \item Symmetric Encryption: \texttt{AES} (Advanced Encryption Standard) with \texttt{128}-bit or \texttt{256}-bit keys, or \texttt{ChaCha20} stream cipher
    \item Mode of Operation: \texttt{GCM} (Galois/Counter Mode) or \texttt{CCM} (Counter with CBC-MAC)
    \item Authentication: \texttt{POLY1305} (specifically for ChaCha20) or \texttt{CCM\_8} indicating \texttt{CCM} variant with 8-byte authentication tag
    \item Integrity: \texttt{SHA} (Secure Hash Algorithm) variants (\texttt{SHA256} or \texttt{SHA384})
\end{itemize}

\begin{figure}[ht!]
    \centering  \includegraphics[width=0.7\linewidth]{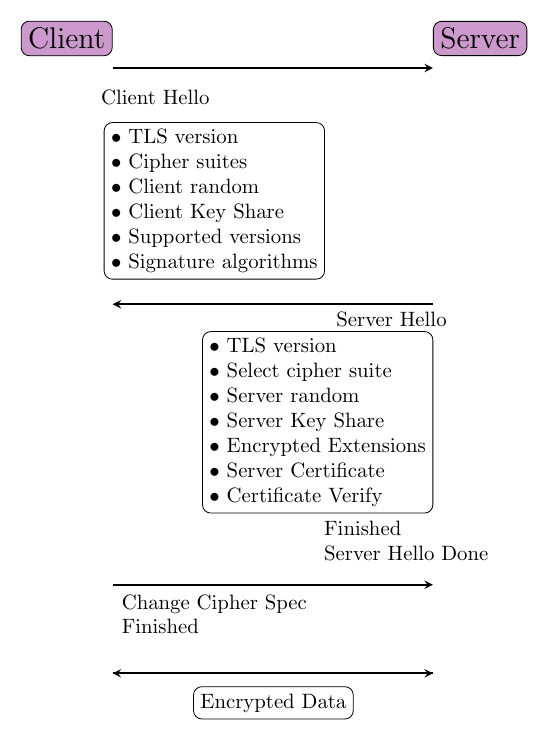}
    \caption{Message flow in the TLS 1.3 handshake}
    \label{fig:TLS-13}
\end{figure}

Moreover in the TLS 1.3 all handshake messages after the ClientHello are now encrypted.
This is a result of the reconstruction of the handshake protocol.
The updated handshake has not only become more secure, but also significantly faster.
In the TLS 1.2 two round trips were needed to establish a secure connection.
Now the Client can put everything Server needs in Client hello shortening the connection to just one round trip.
The message flow in the TLS 1.3 handshake is presented in the Fig~\ref{fig:TLS-13}

\noindent
Most significantly at the moment of receiving Client Hello, a Server has everything required to generate the Master Secret (Client \& Server key shares, Client \& Server Randoms).
Therefore the notion of PMS is not used anymore in TLS 1.3 and after preparation Server Hello message can be encrypted before sending to Client.
Once receiving Server Hello, Client can verify Server's certificate, generate the same MS and send its Change Cipher Spec message for encryption validation.
The HTTPS connection is established one step faster than in TLS 1.2, and having the Server Hello encrypted less information can be stolen by malicious attack. 

\noindent
\subsubsection[Example of full TLS 1.3 handshake]{Example of full TLS 1.3 handshake\protect\footnote{See \texttt{tls13.xargs.org} for key exchange generation using X25519 and further details}}

\vspace{5pt}
\noindent
\textbf{Client Hello}
\begin{enumerate}
    \item Record Header, Handshake Header -- initiate the communication
    \item Client Version: TLS 1.2 -- always 1.2 in the beginning to prevent the session from being terminated by a middlebox that does not recognize versions of the protocol other than 1.2
    \item Client Random: 32 bytes of random data
    \item Session ID: random data -- for compatibility with TLS 1.2
    \item Client Cipher Suites: \texttt{TLS\_AES\_256\_GCM\_SHA384},\\ \texttt{TLS\_CHACHA20\_POLY1305\_SHA256} -- ordered list of supported cipher suites in the order preferred by the Client
    \item Compression Methods: for compatibility with TLS 1.2
    \item Extensions: 
    \begin{itemize}
        \item Extension Length 
        \item Server Name, EC Point Formats
        \item Supported Groups: \texttt{x25519}, \texttt{secp256r1}, \texttt{secp384r1} (supported curves for ECC)
        \item Session Ticket, Encrypt-Then-MAC, Extended Master Secret
        \item Signature Algorithms: \texttt{ECDSA-SECP256r1-SHA256},\\ \texttt{ECDSA-SECP384r1-SHA384}, \texttt{RSA-PSS-PSS-SHA256},\\ \texttt{ED25519}, \dots (signature algorithms supported by Client)
        \item Supported Versions: assign value for TLS 1.3 support
        \item Pre-Shared Key Exchange Modes
        \item Key Share: {\scriptsize\texttt{358072d6365880d1aeea329adf9121383851ed21a28e3b75e965d0d2cd166254}}\\(public key generated with X25519 -- key exchange via curve25519)
        
    \end{itemize}
\end{enumerate}
\vspace{5pt}
\noindent
\textbf{Server Hello}
\begin{enumerate}
    \item Record Header, Handshake Header
    \item Server Version
    \item Server Random: 32 bytes of random data
    \item Session ID: for TLS 1.2 compatibility
    \item Cipher Suite: \texttt{TLS\_AES\_256\_GCM\_SHA384} -- ciphersuite selected from Client's list
    \item Compression Method -- method selected from the list from Client. \texttt{null} in TLS 1.3
    \item Extensions:
    \begin{itemize}
        \item Extension Length
        \item Supported Versions: assign value for TLS 1.3 support
        \item Key Share: {\scriptsize\texttt{9fd7ad6dcff4298dd3f96d5b1b2af910a0535b1488d7f8fabb349a982880b615}}\\(public key generated with X25519 -- key exchange via curve25519)
    \end{itemize}
\end{enumerate}
\textbf{Server Handshake Keys Calc}
\begin{enumerate}
    \item Server secret key: {\scriptsize\texttt{909192939495969798999a9b9c9d9e9fa0a1a2a3a4a5a6a7a8a9aaabacadaeaf}}
    \item Server calculates shared secret multiplying
    \begin{itemize}
        \item Client public key share
        \item Server secret key
    \end{itemize}
    {\scriptsize \texttt{df4a291baa1eb7cfa6934b29b474baad2697e29f1f920dcc77c8a0a088447624}}
    \item Server calculates \texttt{SHA384} hash (hash function from agreed ciphersuite) of Client Hello and Server Hello up to this point without Record Headers.
    \item Using \texttt{HKDF} (H-MAC Key Derivation Functions), secrets are calculated. 

    \begin{itemize}
        \item handshake secret
        \item Server handshake traffic secret
        \item Client handshake traffic secret
        \item Server handshake key
        \item Server handshake IV
        \item Client handshake key
        \item Client handshake IV
    \end{itemize}
\end{enumerate}
\textbf{Client Handshake Keys Calc}
\begin{enumerate}
    \item Client secret key: {\scriptsize\texttt{202122232425262728292a2b2c2d2e2f303132333435363738393a3b3c3d3e3f}}
    \item Client calculates shared secret multiplying
    \begin{itemize}
        \item Client public key share
        \item Server secret key
    \end{itemize}
    {\scriptsize \texttt{df4a291baa1eb7cfa6934b29b474baad2697e29f1f920dcc77c8a0a088447624}}
    \item Client calculates \texttt{SHA384} hash (hash function from agreed ciphersuite) of Client Hello and Server Hello up to this point without Record Headers.
    \item Using \texttt{HKDF} (HMAC-based Key Derivation Functions), secrets are calculated.
    The values of the secrets are the same as these calculated in Server Handshake Keys Calc.
\end{enumerate}
\textbf{From this point the connection is now encrypted}

\subsection{AES-GCM encryption}
AES-GCM (Advanced Encryption Standard - Galois/Counter Mode) is a mode of operation for the AES block cipher which has a widesperad support in both TLS 1.2 and TLS 1.3.
Essentially, in TLS 1.3, AES-GCM is recommended as one of the supported cipher suites due to its security and efficiency properties.
It combines symmetric key encryption with a universal hash function for authenticated encryption and data integrity verification.
AES-GCM uses the AES block cipher in Counter Mode (CTR) for encryption. In CTR mode, AES encrypts a unique counter value for each block of plaintext to produce a stream of ciphertext. 
This operation is reversible, meaning decryption can be performed by encrypting the same counter values.
Alongside encryption, AES-GCM employs Galois field multiplication to provide message authentication. 
This involves multiplying the ciphertext blocks with a unique value derived from the encryption key and the counter value. 
The result of this multiplication is incorporated into the authentication process to ensure the integrity of the data.
After encrypting the plaintext and performing Galois field multiplication, AES-GCM generates an authentication tag, also known as a Message Authentication Code (MAC). 
This tag serves as a cryptographic checksum that is appended to the ciphertext. 
It ensures that the ciphertext has not been tampered with or altered during transmission.

AES-GCM provides both confidentiality (through encryption) and integrity and authenticity (through authentication) in a single operation, making it a popular choice for securing data in various applications, including network communication protocols like TLS. 
Its efficient performance and strong security properties make it well-suited for modern cryptographic applications.

\newpage
\section{TLSNotary}

\subsection{Overview}
The main motivation behind the TLSNotary protocol is to allow the Client to securely and simply confirm the authenticity of its data, obtained in secured connections to the Server, to an external party.
Since this entity is responsible for verifying the provenance of the data, it is called a \textbf{Verifier} in the TLSNotary protocol.
One way this can be accomplished is by having the Verifier and Client connect together over a 3P-TLS connection and then using multi-party computation (MPC) protocols to jointly prepare the proof of authenticity and verify it.
However, since the MPC phase does not reveal anything about plaintext data of the TLS session it is possible to outsource the MPC phase to the another party -- trusted Notary. 
The Notary can sign (notarize) the data without seeing the plaintext, acting as a third party in the 3P-TLS protocol.
Then Client can further redact the notarized data in the selective disclosure phase and finally pass it to the Verifier just for the verification of the notarized data.
This is the second way in which the TLSNotary objective can be achieved.
The data signed that way is reusable (can be shared with multiple Verifiers) and portable as there is no need to share any encrypted or decrypted data with Verifier, just a notarized proof.
The important downside is that the Verifier must absolutely trust the Notary to accept the signed proof.
In both cases the Notary (or Verifier) is not man-in-the-middle.
From the Server's perspective, the connection with the Client is a standard communication over TLS, and the Notary remains transparent to the Server.
Regardless of the choice of scenario, the TLSNotary protocol is divided into three phases

\begin{enumerate}
    \item Client connects with Server over TLS, while Notary (or Verifier) participates in this connection -- jointly operates TLS without seeing data requested by Client in plain text. 
    Notary (or Verifier) joining Client-Server TLS session does not disrupt it and does not lower its security level due to the multi-party computation (MPC) protocols used
    \item Client selectively prove the authenticity of arbitrary parts of the data to the Notary.
    This phase is called selective disclosure (SD).
    Client can redact sensitive data before showing it to Verifier or use ZK Proofs of notarized transcripts to prove properties of the data without revealing the data itself.
    \item Verifier validates the proof from Client and is able to make claims about it in data verification (DV) phase.
    The data origin can be verified inspecting the Server certificate.
\end{enumerate}

\begin{figure}[ht]
    \centering
    \includegraphics[width=\linewidth]{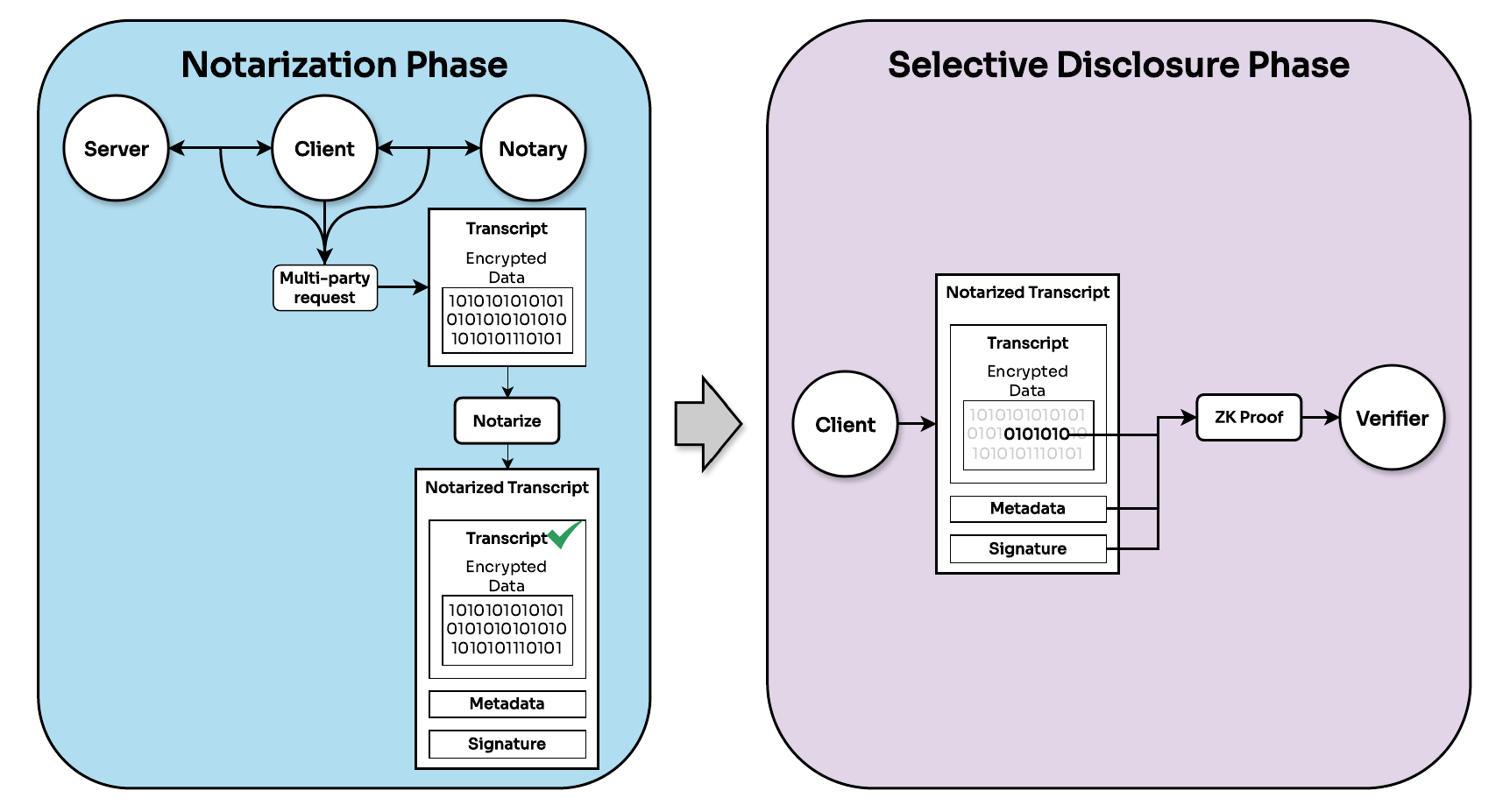}
    \caption{Schematic workflow of the TLSNotary protocol after the Client and Notary connect with the Server using MPC. Diagram used courtesy of TLSNotary Team~\cite{TLSNotary_y2021}, with notation and graphics adjustments by authors of this document.}
    \label{fig:TLSNotary}
\end{figure}
\noindent

\subsection{Use Cases and TLS Support} 
TLSNotary currently supports TLS 1.2. 
TLS 1.3 support is supposed to be added in 2024. 
Some of the use cases of the protocol include
\begin{itemize}
    \item Prove you have access to an account on a web platform
    \item Prove a website showed specific content on a certain date
    \item Prove you have private information about yourself (address, birth date, health, etc.)
    \item Prove you have received a money transfer using your online banking account without revealing your login credentials or sensitive financial information
    \item Prove you received a private message from someone
\end{itemize}

\subsection{Protocol interface}

The TLSNotary protocol operates in three main phases enabling secure and verifiable data exchange between a Client, Server, and Notary. 
This overview provides a detailed examination of each phase: MPC-TLS, Selective Disclosure and Data Verification.
We particularly emphasize the heart of the protocol which is Phase 1: MPC-TLS. 
It includes a subsection \ref{sec:notarization} dedicated to the notarization process when MPC protocol involves the Notary.
We refer to the Notarization as \textit{Phase 1.5} to highlight its integral role in the MPC-TLS process. 

\subsubsection{Phase 1: MPC-TLS} 

\noindent
During the MPC-TLS phase Client and the Notary run an MPC protocol enabling the Client to connect to, and exchange data with, a TLS-enabled Server. 

\subsubsubsection{Secure three party TLS 1.2 handshake}
The first step is TLS handshake to establish TLS connection between Server and Client. 
Client starts the TLS 1.2 handshake, but after initialization all of the cryptographic TLS operations are performed together with the Notary.
Client and the Notary construct their shared TLS session key in such a way they both never learn the full key.
In standard TLS connection between Server and Client the session key is established by creating Pre-Master Secret (PMS) which is computed within ECDH Protocol (PMS is the $x$ coordinate of a point on the EC).
In TLSNotary Client and Notary compute additive shares of PMS which at the end of the key exchange is delivered  to the Server.
3P-TLS key exchange is performed using 3-party ECDH protocol in the following way
\begin{enumerate}
    \item Server sends his public key $S_{pk}$ to the Client
    \item Client forwards Server's public key $S_{pk}$ to the Notary
    \item Client generates random key share $C_{sk}$ and creates his public key share $C_{pk} = C_{sk} \cdot G$
    \item Notary generates his random key share $N_{sk}$ and creates his public key share $N_{pk} = N_{sk} \cdot G$ 
    \item Notary sends his public key share $N_{pk}$ to Client 
    \item Client calculates public key for the TLS session $P_{pk} = C_{pk} + N_{pk}$ and sends it over to the Server
    \item Client calculates EC point $C = (x_C, y_C) = C_{sk} \cdot S_{pk}$
    \item Notary calculates EC point $N = (x_N, y_N) = N_{sk} \cdot S_{pk}$
    \item PMS is given by $x$ coordinate of EC point $R = C+N$. It is given by 
    \begin{equation}
        F_x(R) = \left( \frac{y_N - y_C}{x_N - x_C} \right)^2
        - x_C - x_N,
        \label{eq:TLSNotary-PMS}
    \end{equation}
    where $F_x(\cdot)$ is the function which returns $x$ coordinate of the EC point.
    In order to calculate $x$ using formula (\ref{eq:TLSNotary-PMS}) Client and Notary would have needed to share their secrets, which can not happen.
    {
    To calculate $F_x(R)$ securely (so Client and Notary do not share their secrets) protocols OLE flavors of the \texttt{A2M}~\cite{KOS16} i \texttt{M2A}~\cite{XYW23} are used}
    \begin{itemize}
    \item First using \texttt{A2M}$(y_N + (-y_C)) = A_N \cdot A_C$ and \texttt{A2M}$(x_N + (-x_C)) = B_N \cdot B_C$ Client and Notary convert their additive shares to multiplicative ones.
    Then the formula (\ref{eq:TLSNotary-PMS}) translates to
    \begin{equation}
        F_x(R) = C_N \cdot C_C - x_C - x_N
    \end{equation}
    where $C_N = \left( A_N/B_N \right)^2$ oraz $C_N = \left( A_C/B_C \right)^2$
    \item Using \texttt{M2A}$(C_N \cdot C_C) = D_N + D_C$ Client and Notary convert their multiplicative shares to additive.
    \item 
    Finally, PMS is given by the formula
    \begin{equation}
        \texttt{PMS} = \left( D_N - x_N \right)
        - \left( D_C - x_C \right)
    \end{equation}    
    which can be calculated by Client and the Notary without revealing their secret key-shares $C_{sk}$ and $N_{sk}$ respectively
    \end{itemize}
    \item Server calculates PMS on its own following standard  ECDH key exchange procedure and taking $x$ coordinate of the shared point as a
    \begin{equation}
        \texttt{PMS} = F_x(P_{pk} \cdot S_{sk}),
    \end{equation}
    where $S_{sk}$ is Server secret key
\end{enumerate}

\noindent
It is worth noting that the PMS obtained by Server and the Client/Notary is identical since $P_pk = C_{sk} \cdot G + N_{sk} \cdot G$ and $S_{pk} = S_{sk} \cdot G$.
Thus the following equality holds
\begin{equation}
\begin{array}{lll}
    \texttt{PMS}_{C/N} &=& F_x(C_{sk} \cdot S_{pk} + N_{sk} \cdot S_{pk}) \\
    &=& F_x(C_{sk} \cdot S_{sk} \cdot G + N_{sk} \cdot S_{sk} \cdot G)\\
    &=& F_x((C_{sk} + N_{sk})\cdot G \cdot S_{sk}) \\
    &=&  F_x(P_{pk}\cdot S_{sk}) \\
    &=& \texttt{PMS}_S
\end{array}
\end{equation}
and the key exchange is correct.
{
To make sure that OLE based \texttt{A2M} and \texttt{M2A} protocols are performed correctly and securely steps 9 and 10 are performed twice -- with Client and Notary switching roles from OLE Sender to Receiver.
Then an equality check protocol with garbled circuits is performed to validate if \texttt{PMS}s calculated in both takes are the same.
Even though this approach might be slightly less efficient compared to~\cite{XYW23}, it is more conservative and has strong security guarantees.
}

Once the key exchange protocol is complete, the Client is ready to exchange encrypted data with the Server. 
All of the Client's plain text data is subject to joint encryption with Notary. 
Similarly, the decryption of the Server data and the MAC computation for the ciphertext is performed jointly by the Client and the Notary, however, in such a way that the Notary never sees the plain text.
The MPC protocols that allow such cooperation are \texttt{DEAP}, used for data encryption and decryption, and the \texttt{2PC MAC} used to calculate the MAC for the cipher text

\vspace{10pt}
\noindent\fbox{
    \parbox{\textwidth}{
    \noindent
    \textbf{Encryption scheme}

    \noindent
    Client inputs his plaintext as a private input.
    Client and Notary input their TLS key shares as inputs to \texttt{DEAP}.
    Output message is encrypted and together with MAC it's sent to the Server.

    \vspace{5pt}
    \noindent
    \textbf{MAC}

    \noindent
    Message Authentication Code (MAC) is used to authenticate the ciphertext and check if it has not been tampered during transfer.
    In \texttt{AES-GCM} cipher suites (compatible with TLS 1.2 and 1.3) MAC is calculated using \texttt{GCTR} and \texttt{GHASH} functions.
    Client and Notary calculate MAC using secure \texttt{2PC MAC} protocol, which ensures compatibility wit TLS and privacy of the inputs. 
    The parties only know their own \texttt{GHASH} and \texttt{GCTR} shares and use them to compute MAC jointly.

    \vspace{5pt}
    \noindent
    \textbf{Decryption scheme}

    \noindent
    Client receives ciphertext and its MAC from the Server.
    First, Client and Notary verify the MAC to authenticate the ciphertext by running \texttt{2PC MAC} protocol.
    Client and Notary input their TLS key shares as inputs to \texttt{DEAP}.
    Resulting plaintext is visible only to Client.
}
}

\subsubsubsection{Dual Execution with Asymmetric Privacy}
Dual Execution with Asymmetric Privacy (\texttt{DEAP}) protocol, being used in the encryption and decryption schemes, is the heart of phase one of TLSNotary.
The protocol takes place between Client and Notary.
After execution Client preserves full privacy -- none of his inputs are exposed.
Although the Notary is obligated to reveal his private inputs (TLS key-share) after the TLS connection with the Server is over.
Hence the name -- privacy-wise the protocol is asymmetric.
The protocol comes down to computation of values of the function $f(x_C,x_N) = v$, where $x_C$ and $x_N$ are the private inputs of the Client and the Notary. 
In the last step of the protocol $x_N$ is being revealed.

\vspace{10pt}
\noindent\fbox{
    \parbox{\textwidth}{
\noindent
The meaning of $f(x_C,x_N) = v$ variables with respect to the TLSNotary encryption and decryption schemes is the following:
\begin{itemize}
    \item $x_N$ always consists of Notary's TLS key share ($x_N = k_N$). 
    \item $x_C$ can contain Client's key share and plaintext or ciphertext.
Plaintext when encrypting data being sent to Server and ciphertext received from the Server for the decrytpion scheme.
\item $v$ is the resulting ciphertext being sent to the Server in the encryption procedure.
In the decryption procedure $v$ is the plaintext decrypted from Server's message
\end{itemize}
}}
\vspace{10pt}

\noindent
\texttt{DEAP} consists of three phases: (1) Setup, (2) Execution and (3) Equality Check.
Client and the Notary perform the first two steps of the protocol, but Notary waits to perform step three until the TLS connection with Server is closed so as not to reveal its private input before the end of the session.

\begin{figure}[ht!]
    \centering
    \includegraphics[width=0.9\linewidth]{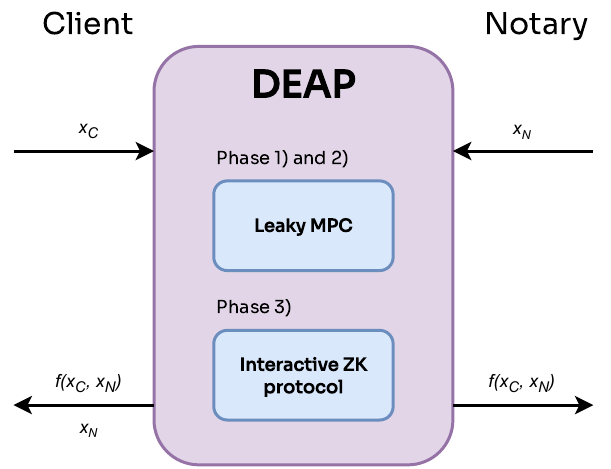}
    \caption{Black-box diagram of \texttt{DEAP} protocol}
    \label{fig:DEAP-blackbox}
\end{figure}
\noindent
{
Figure~\ref{fig:DEAP-blackbox} illustrates the overview of \texttt{DEAP}  protocol. 
The protocol comprises two primary elements:
\begin{itemize}
    \item Leaky Multi-Party Computation (Phases 1 and 2): Employs multi-party computation to process inputs $x_C$ and $x_N$ from the Client and Notary, respectively.
    \item Interactive Zero-Knowledge Protocol (Phase 3): Implements a zero knowledge protocol via MPC techniques, specifically garbled circuits~\cite{Jawurek_2013}, to achieve malicious security for the protocol.
\end{itemize}

\texttt{DEAP} ensures security against malicious behavior from both parties while providing asymmetric privacy. 
The Client's input remains private throughout, but the Notary must reveal $x_N$ to the Client during the Equality Check in Phase 3. 
This asymmetry is a necessary trade-off to maintain protocol security through the interactive zero-knowledge (IZK) component. 
Noteworthy, the proven malicious security of \texttt{DEAP} implies that any 2-party protocol with asymmetric privacy designed in this manner (Leaky MPC + IZK) will be maliciously secure. 
Thus, such structure not only guarantees robust security while preserving the Client's privacy but also provides a framework for future optimizations in asymmetric privacy protocols.
Potential future directions in optimizing leaky MPC phases, which are generally less complex than optimizing maliciously secure MPC, could include techniques optimizing garbled circuits like~\cite{RR21} or oblivious transfer protocols~\cite{KOS16, YWLZW20}.
Whereas IZK performance can be enhanced using approaches like Quicksilver~\cite{YSWW21} or other MPC IZK protocols \cite{FNO14, CDGORRSZ17, HK20}. 

}

{
We now proceed to a detailed examination of \texttt{DEAP}'s three phases: Setup, Execution, and Equality Check. 
This analysis will elucidate the roles of the Client and Notary, the cryptographic techniques employed, and the security measures implemented in each phase. 
By dissecting these components, we aim to demonstrate how \texttt{DEAP} achieves malicious security and asymmetric privacy.}

\begin{enumerate}
    \item Setup phase
    \item[~]\textit{Notation:}  
    \begin{itemize}
        \item[$\circ$] $e_i(x)$, where $i\in\{C, N\}$ denotes an encoding of message $x$ in the sense of garbled circuit labeling\footnote{See section 2.4.1 Garbling scheme formalization}: $$\texttt{Encode}(x, e_i(x)) = [x]_i.$$
        \item[$\circ$] $[x_C]_i$, where $i\in\{C, N\}$ denotes active input labels of Client's message encoded with Client or Notary's encoding information $e_i(x_C)$ respectively
        \item[$\circ$] $[x_N]_i$, where $i\in\{C, N\}$ denotes active input labels of Notary's message encoded with Client or Notary's encoding information $e_i(x_N)$ respectively
        \item[$\circ$] $d_i$ denotes Client/Notary's decoding information used to decode respective encoding:
        $$\texttt{Decode}(d_i, [x]_i) = x$$
    \end{itemize}

    \item[~]{Setup phase consists of the following steps}  
    \begin{itemize}
        \item Client creates garbled circuit $G_C$ for the function $f(x_C, x_N) = v$ with input labels encoding $\left(e_C(x_C), e_C(x_N)\right)$ and output label encoding $e_C(v)$. 
        Additionally Client creates a binding commitment to the encoding of $v$: $\texttt{com}_{e_C(v)}$ 
        \item Notary creates a garbled circuit $G_N$ for the function $f$ with his corresponding input labels encoding $\left(e_N(x_C), e_N(x_N)\right)$ and output label encoding $e_N(v)$.
        \item Notary generates random-tape for committed Oblivious Transfers using seed $\rho$ and pseudo-random generator \texttt{PRG}$(\rho)$
        \item Notary sends binding commitment to $\rho$ \texttt{com}$_{\rho}$ to Client
        \item Using OT Client gets his active input labels $[x_C]_N$ from Notary
        \item Using OT Notary gets his active input labels $[x_N]_C$ from Client
        \item Client sends his garbled circuit $G_C$ with active input labels $[x_C]_C$ and 
        $\texttt{com}_{e_C(V)}$ to Notary together with $d_C$, which is output decoding information (translation table)
        \item Notary sends his garbled circuit $G_N$ together with his active input labels $[x_N]_N$ and output decoding information $d_N$ to Client
    \end{itemize}
    In conclusion, the setup phase of the \texttt{DEAP} protocol is to create garbled circuits by Client and Notary and prepare for their evaluation.
    Conceptually, this is the execution of the standard first three steps in the garbled circuit protocol where Client is the evaluator for Notary's circuit and vice versa.
    The difference between this scheme and the one presented in Section 2.4 lies primarily in the OT protocol used, using binding commitments
    
    \item Execution phase (executed parallely by Client and Notary)

    \noindent
    This phase introduces some additional notations for the output of the function being evaluated using the \texttt{DEAP} protocol: the variable $v$.
    The symbol $[v]_{N/C}$ denotes output $v$ encoded using Notary's or the Client's encoding information respectively $$[v]_{N/C} = \texttt{Encode}(v, e_{N/C}(v)).$$
    Consequently symbol $v^{N/C}$ denotes output $v$ decoded with Notary's or Client's decoding information $d_{N/C}$ $$v^{N/C} = \texttt{Decode}(d_{N/C}, [v]_{N/C}).$$ 
    Of course, ideally $v^N = v_C = v$ however, this assumption cannot be made at this stage, and verifying it is the goal of the next phases of the \texttt{DEAP} protocol
    \item[~]
    \item[~]{Execution phase consists of the following steps performed in parallel by Client and Notary:}  
    \begin{itemize}
        \item[~] \textbf{Client}:
        \item Evaluates garbled circuit $G_N$ using $[x_N]_N$ and $[x_C]_N$ to acquire $[v]_N$
        \item Decodes $[v]_N$ using $d_N$ to acquire $v^N$
        \item Computes secure hash function $\texttt{H}([v^N]_C, [v]_N)$ which is called $\texttt{check}_C$
        \item Computes a commitment with a private secret key $r$\\ $C(\texttt{check}_C, r)=\texttt{com}_{\texttt{check}_C}$
        \item Client sends \texttt{com}$_{\texttt{check}_N}$ to Notary
        \item Waits to receive $[v]_C$ from Notary
        \item Check if $[v]_C$ is authentic, aborting if not. 
        \item Client decodes $[v]_C$ using decoding information $d_C$ \\\texttt{De}$(d_C, [v]_C) = v^C$)
        \item[~] \textit{Note:} At this stage, if Notary is malicious, Client could detect that $v^C \neq v^N$.
        However, Client must not react in this case. 
        He proceeds with the protocol regardless, having his authentic output $v^C$
        \item[~] \textbf{Notary:}
        \item Evaluates garbled circuit $G_C$ using provided $[x_C]_C$ and $[x_N]_C$ to acquire $[v]_C$
        \item Checks $[v]_C$ against the commitment \texttt{com}$_{e_C(v)}$ sent by Client in the Setup.
        Notary aborts if they do not match
        \item Decodes $[v]_C$ to $v^C$ using $d_C$
        \item Computes secure hash function \texttt{H}$([v]_C, [v^C]_N)$ which is called $\texttt{check}_N$ and stores it for the equality check later
        \item Sends $[v]_C$ to Client
        \item Receives $\texttt{com}_{\texttt{check}_C}$ from Client and stores it for equality check
    \end{itemize}
    \item Equality check phase:
    \begin{itemize}
        \item Notary opens his garbled circuit $G_N$ and opens committed oblivious transfer by sending $x_N$, $\rho$ and global offset of his garbled circuit $\Delta_N$ to Client
        \item Client can now derive suspected input labels to Notary's garbled circuit $\left(e_N(x_C)^{\ast}, e_N(x_N)^{\ast}\right)$
        \item Client uses $\rho$ to open all of Notary's oblivious transfers for $[x_C]_N$ and verifies that they were performed honestly.
        Otherwise aborts
        \item Client verifies that $G_N$ was garbled honestly by checking that garbling of Notary input labels he derived $\left(e_N(x_C)^{\ast}, e_N(x_N)^{\ast}\right)$ gives garbled circuit $G_N$.
        Otherwise aborts
        \item Client opens $\texttt{com}_{\texttt{check}_C}$ by sending \texttt{check}$_C$ and $r$ to Notary
        \item Notary verifies $\texttt{com}_{\texttt{check}_C}$ and asserts $\texttt{check}_C == \texttt{check}_N$.
        Otherwise aborts
    \end{itemize}
\end{enumerate}
After evaluation {Notary is convinced} $v^A$ is correct ie $f(x_C,x_N) = v^A$ without knowing the value of $x_C$. 
Security issues of that protocol related to cases of Malicious Client and Malicious Notary are further investigated in~\cite{Jawurek_2013, Dupin_2018}\footnote{In the cited works TLSNotary Client is represented by Alice, and Notary is represented by Bob}.

\subsubsubsection{Encryption protocol}

\noindent
Using \texttt{DEAP} protocol, without any alteration, two parties encrypt the message using a {block cipher in counter-mode}, standard mode of AES encryption in TLS. 
The encryption protocol reads as follows:
\begin{enumerate}
    \item The Client and Notary directly compute the ciphertext for each block of a message the Client wishes to send to the Server
    \begin{equation}
        f(k_C, k_N, ctr, p) = \texttt{Enc}(k_C \oplus k_N, ctr)\oplus p = c,
    \end{equation}
    where $k_C$ and $k_N$ are Client and Notary cipher key shares, $ctr$ is the \textit{counter block}, $p$ is the block of the plaintext, $c$ is the corresponding block of ciphertext encrypted with AES/GCM ($c = \texttt{Enc}(k, ctr) \oplus p $)
    \item The Client creates a commitment to the plaintext active labels $\texttt{Com}([p]_N, r)$ for the Notary's garbled circuit $G_N$, where $r$ is a random key known only to the Client
    \item Client sends $\texttt{Com}([p]_N, r)$ to be used in the \texttt{DEAP} protocol later
    \item Client commits to $[p]_N$ {before} Notary reveals his $\Delta_N$ in the equality check phase of \texttt{DEAP}
    \item If $\texttt{Com}([p]_N, r)$ is a commitment to valid labels, then it must be a valid commitment to the plaintext $p$ as learning the complementary wire label for any bit of $p$ prior to learning $\Delta_N$ is impossible
\end{enumerate}
Both parties see the resulting ciphertext and execute the 2PC MAC protocol to compute the Message Authentication Code (MAC) for the ciphertext.
The Client then dispatches the ciphertext and the MAC to the Server.

\subsubsubsection{Secure \texttt{2PC MAC} protocol}

\texttt{AES-GCM} (Advanced Encryption Standard - Galois Counter Mode) is a mode of operation for the block cipher supported both by TLS 1.2 and 1.3. 
Although it is more commonly used in TLS 1.3 as it is recommended due to its security and efficiency properties.
The \texttt{MAC} (Message Authentication Code) in \texttt{AES-GCM} is computed using a Galois/Counter mode technique, which combines symmetric key encryption (\texttt{AES} in Counter Mode) with a universal hash function. 
\texttt{AES-GCM} uses \texttt{GCTR} function for encryption and \texttt{GHASH} function as a universal hash.
The authentication tag (\texttt{MAC}) is then computed by XORing the output of the \texttt{GHASH} function with the \texttt{GCTR} encryption of a special nonce called the authentication tag nonce, derived from the encryption key. 
This XOR operation helps ensure that even if an attacker can modify the ciphertext, they cannot easily forge a valid authentication tag without knowing the encryption key.

\noindent
The goal of \texttt{2PC MAC} protocol is to compute MAC in such a way that neither Client nor Notary learns the other party \texttt{GHASH} key-share $H_{C/N}$.

\vspace{10pt}
\noindent\fbox{
    \parbox{\textwidth}{
\noindent
The inputs of \texttt{2PC MAC} for each party are following:

\noindent
Client has
\begin{itemize}
    \item ciphertext blocks $X_1, X_2, \dots, X_n$.
    \item \texttt{GHASH} key share $H_C$
    \item Share of \texttt{GCTR} output encrypted with $k_C$ key, denoted $\texttt{GCTR}_C$
\end{itemize}

\noindent
Notary has
\begin{itemize}
    \item ciphertext blocks $X_1, X_2, \dots, X_n$.
    \item \texttt{GHASH} key share $H_N$
    \item Share of \texttt{GCTR} output encrypted with $k_N$ key, denoted $\texttt{GCTR}_N$
\end{itemize}
}}
\vspace{10pt}

\noindent
For the single ciphertext block $X_1$ the MAC computation consists of three steps
\begin{enumerate}
    \item Client calculates $X_1 \cdot H_C \oplus \texttt{GCTR}_C = MAC_C$
    \item Notary calculates $X_1 \cdot H_N \oplus \texttt{GCTR}_N = MAC_N$
    \item Notary sends $MAC_N$ to Client who calculates $MAC = MAC_N \oplus MAC_C$

\noindent
Please note that  $"\cdot"$ denotes a multiplication in a finite field, as described in \texttt{GCM} specification~\cite{Dworkin2007} 
\end{enumerate}

\noindent
The situation is more complicated for an arbitrary long ciphertext, which is divided into $n$ blocks.
Note that both parties hold their {additive} \texttt{GHASH} keyshares $H_C$ and $H_N$, namely $H = H_N \oplus H_C$\footnote{All of the operations within Galois/Counter Mode happen in the extension field ${GF}(2^{128})$, where by addition is defined as XOR and multiplication in finite field is described in~\cite{Dworkin2007}}.
The calculation of \texttt{GHASH} function secured with \texttt{GHASH} key $H$ for $n$ blocks of ciphertext is then given by the following formula
\begin{equation}
    \texttt{GHASH}(X_1, \dots, X_n; H) = X_1\cdot H^n \oplus X_2\cdot H^{n-1} \oplus \dots \oplus X_{n}\cdot H.
    \label{eq:ghash}
\end{equation}
\noindent
The powers of $H$ are regarded with respect to finite field multiplication "$\cdot$", thus $H_N^n \oplus H_C^n \neq (H_N \oplus H_C)^n$.
Because of that the $n$-th powers of respective shares $H_C$ and $H_N$ cannot be computed  locally.
{
Although, as we operate in the extension field ${GF}(2^{128})$ an operation known as "free squaring" holds, namely $H_N^2 \oplus H_C^2 = (H_N \oplus H_C)^2$.
This allows to reduce the bandwidth for computing the higher powers of $H$ by a factor of two.
}
To compute higher powers of $H$ the following protocol for calculating \texttt{2PC GHASH} function is used

\vspace{10pt}
\noindent\fbox{
    \parbox{\textwidth}{
\noindent
\texttt{2PC GHASH} protocol interface:

\noindent
\textbf{Inputs}: Client holds \texttt{GHASH} keyshare $H_C$, Notary holds $\texttt{GHASH}$ keyshare $H_N$ such that $H_N \oplus H_C = H$ \\
Both Client and Notary hold ciphertext divided into $n$ blocks: $X_1, X_2, \dots, X_n$

\noindent
The procedure to calculate \texttt{GHASH} function given by Eq.(\ref{eq:ghash}) is the follwing
\begin{enumerate}
    \item Using \texttt{A2M} protocol both parties convert additive keyshares to  multiplicative \texttt{A2M}$(H_N, H_C) = (H_N^{\ast}, H_C^{\ast})$ such that $H = H_N^{\ast}\cdot H_C^{\ast}$
    \item Each party calculates $n$ powers of their multiplicative share
    \begin{itemize}
        \item Client evaluates $(H_C^{\ast})^2, (H_C^{\ast})^3, \dots, (H_C^{\ast})^n$
        \item Notary evaluates $(H_N^{\ast})^2, (H_N^{\ast})^3, \dots, (H_N^{\ast})^n$
        \item Such that e.g. $(H_C^{\ast})^2 \cdot (H_N^{\ast})^2 = H^2$
    \end{itemize}
    \item Both parties use \texttt{M2A} protocol $n$ times to convert multiplicative shares back to additive for each power, e.g. $\texttt{M2A}((H_N^{\ast})^2, (H_C^{\ast})^2) = (H_N^{2}, H_C^{2})$ such that $H_N^{2} \oplus H_C^{2} = H^2$
    \item Each party can locally compute their \texttt{GHASH} function share, namely
    \begin{itemize}
        \item Client evaluates $\texttt{GHASH}_C = X_1\cdot H_C^n \oplus X_2\cdot H_C^{n-1} \oplus \dots \oplus X_{n}\cdot H_C$
        \item Notary evaluates $\texttt{GHASH}_N =X_1\cdot H_N^n \oplus X_2\cdot H_N^{n-1} \oplus \dots \oplus X_{n}\cdot H_N$
    \end{itemize}
\end{enumerate}
}}
\vspace{10pt}

\noindent
Having computed respective \texttt{GHASH}$_{C/N}$ function shares, Client and Notary can XOR them with respective \texttt{GCTR}$_{C/N}$ shares to obtain additive $MAC_{C/N}$ shares, and finally combine them into $MAC$ for the given ciphertext.
Note that $\texttt{A2M}$ and $\texttt{M2A}$ protocols used for \texttt{GHASH}$_{C/N}$ calculation are implemented with oblivious linear evaluation using techniques introduced in the works~\cite{KOS16, XYW23} robustly adjusted by TLSNotary~\cite{TLSNotary_y2021}.

\subsubsubsection{Decryption protocol}

\noindent
The decryption protocol is similar to the \texttt{DEAP} used for encryption.
The difference is that \texttt{DEAP} is used for every block of ciphertext to compute the {masked encrypted counter block}

\begin{enumerate}
    \item Client chooses a mask $z$ which hides the encrypted message from the Notary.
    The \texttt{DEAP} function reads as follows
    \begin{equation}
        f(k_C, k_N, ctr, p) = \texttt{Enc}(k_C \oplus k_N, ctr) \oplus z = ectr_z
    \end{equation}
    Client can simply remove this mask in order to compute the plaintext
    \begin{equation}
        p = c \oplus ectr_z \oplus z
    \end{equation}
    \item This way Client can retrive the wire labels $[p]_N$ from the Notary using oblivious transfer
    \item As in encryption procedure, the Client creates a commitment $\texttt{Com}([p]_N, r)$ where $r$ is a random key known only to Client
    \item The Client sends this commitment to the Notary to be used in the \texttt{DEAP} protocol later.
\end{enumerate}
The Client has to prove that the labels $[p]_N$ actually correspond to the ciphertext $c$ sent by the Server.
This can be done in one execution using {zero-knowledge} protocol described in~\cite{Jawurek_2013}.
The Notary garbles a circuit $G_N$ which computes
\begin{equation}
    p \oplus ectr = c
\end{equation}
Notice that the Client and Notary will already have computed $ectr$ when they computed $ectr_z$ earlier. 
Conveniently, the Notary can re-use the garbled labels $[ectr]_N$ as input labels for this circuit. For more details on the reuse of garbled labels see~\cite{Afshar_2017}.

\noindent
\subsubsection{Phase 1.5: Notarization\label{sec:notarization}} 

\noindent
While the Client has the capability to directly prove data provenance to the Verifier (by connecting Verifier directly into 2PC-TLS session), there are use cases where it could be advantageous for the Verifier to delegate the verification of the TLS session to the Notary, obtaining a Notarized version of the data as a result.
In this scenario, the Verifier benefits from the notarized data requiring less storage space and alleviating the need for direct participation in the TLS session.
As the part of the TLSNotary protocol, using \texttt{DEAP} to encrypt the data the Client creates authenticated commitments to the plaintext.
The Notary sign them without ever seeing the plaintext. 
This provides the Client with a method to selectively prove the authenticity of specific parts of the plaintext to a application specific Verifier at a later stage.
The high-level idea is that the Client creates a commitment to the active plaintext encoding from the MPC protocol (\texttt{DEAP}) used for Encryption and Decryption.
Client privacy is additional preserved by Client comminting to the Merkle tree of commitments. 
This way the amount of commitments remain hidden.

Additional the Notary signs an artifact known as a Session Header, thereby attesting to the authenticity of the plaintext from a TLS session. 
A Session Header contains a Client's commitment to the plaintext and a Client's commitment to TLS-specific data which uniquely identifies the Server.
The Client can later use the signed Session Header to prove data provenance to an application-specific Verifier.
It's important to highlight that throughout the entire TLSNotary protocol, including this signing stage, the Notary does not gain knowledge of either the plaintext or the identity of the Server with which the Client communicated.

\subsubsubsection{Binding commitment scheme}

Client and the Notary use \texttt{DEAP} protocol to jointly encrypt and decrypt the data in the TLS session.
During setup phase of \texttt{DEAP} Notary creates garbled circuit and sends it to the Client.
The circuit includes Notary's active encoding of the Client's private variables.
In the case of encryption scheme these are encoding of the Client's TLS key-share $e_N(k_C)$ and the plaintext $e_N(p)$.
For the decryption scheme in the setup phase of \texttt{DEAP} Notary again sends the active encoding of Client's TLS key-share and active encoding of $z$ -- private mask for ciphertext selected by Client.
After execution phase of \texttt{DEAP} in the decryption process Client is able to compute the plaintext XORing masked ciphertext with his private mask.
Then he is able to retrieve Notary's active encoding of the plaintext $e_N(p)$ using oblivious transfer. 

In both encryption and decryption scheme Client selects a random key $r$ and commits to the Notary's plaintext encoding.
The commitment $C(e_N(p), r)$ is sent to the Notary before the equality check phase, where Notary reveals his private inputs.
The commitment is used in authdecode protocol later.
Such commitment scheme is more lightweight than standard commitmentsin MPC (e.g. BLAKE3 hash).
Additional by adding the commitments to the Merkle tree of commitments the amount of commitments remains hidden.

For an illustrative example let us suppose we have 8-bit long plaintext message reading $m =$ \texttt{01101000}.
The Notary has created a random encoding for the plaintext consisting of 16 random strings (2 per byte): $e_N = \{w_i^0, w_i^1\}$, where $i = 0, 1, \dots, 7$ indicate byte position in the plaintext message.
Client gets to know active encoding of plaintext, which for the message $m$ reads as follows $e_N(m) = w_0^0 w_1^1 w_2^1 w_3^0 w_4^1 w_5^0 w_6^0 w_7^0$.
Client hashes $e_N(m)$ and commits $\texttt{H}(e_N(m))$ in the Merkle tree.
Now let's suppose Client want to show different data to the Verifier who has received a Notary signature.
The purported data read as follows $m_p =$ \texttt{011010{\color{purple}1}0} (6th bit has been flipped).
Using Notary's signature Verifier can retrieve the full encoding (not only active encoding for message $m$, as Client could) and  select the active encoding for the purported data which reads as follows $e_N(m_p) = w_0^0 w_1^1 w_2^1 w_3^0 w_4^1 w_5^0 {\color{purple}w_6^1} w_7^0$.
Then the hash of the $e_N(m_p)$ can not be found  in the Merkle tree, thus Verifiers know the message $m_p$ has not been notarized.

\noindent
\subsubsection{Phase 2: Selective Disclosure}

\noindent
Client is allowed to selectively disclose the data.
Selective disclosure in the simplest case may be achieved by Client redacting the plaintext.
More generally, Client should be able to use zero knowledge proofs to make certain claims about redacted data.
The proposed upgrades to the Selective Disclosure phase of the TLSNotary protocol include providing higher levels of abstraction to simplify the implementation of proving and verification flows for application developers. 
These upgrades will allow developers to write zero-knowledge (ZK) programs using general purpose programming languages, enhancing both accessibility and flexibility. 
Additionally, specialized functionalities will be introduced to address common challenges, such as handling HTTP and JSON data, thereby streamlining the integration and functionality of the selective disclosure mechanism.

\noindent
\subsubsection{Phase 3: Data Verification}

\noindent
To prove data provenance to a third-party Verifier, the Client provides the following information

\begin{itemize}
    \item Session Header signed by the Notary
    \item Opening information for plaintext commitment
    \item TLS-specific data for Server identification
    \item Identity of the Server
\end{itemize}

\noindent
Then Verifier performs the following verification steps

\begin{itemize}
    \item Verifies that opening information corresponds to the commitment from the Session Header
    \item Verifies that TLS-specific data corresponds to the commitment from the Session Header
    \item Verifies the identity of the Server against the TLS-specific data
\end{itemize}

Next, the Verifier parses the opening with an application-specific parser (e.g. HTTP or JSON) to get the final output. 
Since the Client is allowed to selectively disclose the data, that data which was not disclosed by the Client will appear to the Verifier as redacted.

\section{Summary}
{
This paper presents a comprehensive review of the TLSNotary protocol, a novel approach for providing proof of data authenticity in TLS sessions without compromising privacy or security. 
The protocol leverages multi-party computation (MPC) techniques, particularly the \texttt{DEAP} (Dual Execution with Asymmetric Privacy) protocol, to enable a Client to prove the provenance of data obtained from a TLS session to a third-party Verifier. 
The paper details the cryptographic primitives underlying TLSNotary, provides an overview of TLS, and thoroughly examines the three phases of the TLSNotary protocol: MPC-TLS with Notarization, Selective Disclosure and Data Verification.

The analysis reveals that TLSNotary offers a robust solution for secure and private data attestation, addressing a critical need in today's digital landscape. 
However, the protocol's complexity and the rapidly evolving nature of cryptographic standards present both challenges and opportunities for further research and development.

As the field of secure computation continues to advance, it becomes increasingly important to contextualize TLSNotary within the broader ecosystem of TLS attestation frameworks. 
Future research should therefore focus on comparative analyses of various approaches, including other MPC-TLS architectures (e.g., zkPass, DECO, PADO), proxy architectures (e.g., Reclaim, Janus), and Trusted Execution Environments (e.g., Clique). 
Such studies would provide valuable insights into the relative strengths and limitations of each method, helping to guide future developments in the field.

Moreover the paper's detailed examination of \texttt{DEAP} highlights its potential for optimization both in the areas of Leaky MPC and Interactive ZK protocols. 
This, coupled with the need to extend TLSNotary to support evolving standards like TLS 1.3, opens up significant avenues for performance enhancements and expanded applicability.

Finally, while the paper demonstrates the theoretical soundness of TLSNotary, practical implementation and real-world adoption remain areas ripe for exploration. 
Future practical work should focus on developing more efficient selective disclosure mechanisms, exploring applications, and conducting thorough analyses of scalability and performance in comparison to other attestation frameworks.
}
\section{Acknowledgments}
We are deeply grateful to the TLS Notary team members, particularly Thomas (th4s), Sinu, and Hendrik, for their comprehensive review and proofreading, insightful discussions, and valuable contributions.
Furthermore, we would like to express our gratitude to Mike Rosulek from Oregon State University for providing a response to a simple yet fundamental question regarding the security of garbled circuits back in March 2024.
Additionally, we would like to acknowledge the engineers at vlayer for their comprehensive proofreading of the article and detailed questions, which enhanced the clarity of the text.

\newpage
\bibliography{3PTLS}

\end{document}